\documentclass[aps,prd,twocolumn,groupedaddress,showpacs]{revtex4}
\usepackage{amsmath}
\usepackage{amssymb}
\usepackage{graphicx}
\begin{document}
%

\title{Quintessence with a constant equation of state
in hyperbolic universes}

\author{Ralf Aurich}
\email[]{aurich@physik.uni-ulm.de}

\author{Frank Steiner}
\email[]{steiner@physik.uni-ulm.de}

\affiliation{
Abteilung Theoretische Physik, Universit\"at Ulm,
Albert-Einstein-Allee 11, D-89069 Ulm, Germany
}

\begin{abstract}
Quintessence models leading to a constant equation of state
are studied in hyperbolic universes.
General properties of the quintessence potentials $V(\phi)$
are discussed, and for some special cases also the exact analytic
expressions for these potentials are derived.
It is shown that the observed angular power spectrum of the
cosmic microwave background (CMB) is in excellent agreement with
some of the quintessence models even in cases with negative curvature.
It is emphasized that due to a
$(w_\phi, \Omega_\phi, \Omega_{\hbox{\scriptsize c}})$-degeneracy
a universe with negative spatial curvature cannot be excluded.
\end{abstract}


\maketitle

\section{Introduction}

The recent observations of the anisotropy of the cosmic microwave background (CMB)
\cite{Netterfield_et_al_2001,Lee_et_al_2001,Halverson_et_al_2001}
together with the power spectrum of the large scale structure (LSS)
\cite{Peacock_Dodds_1994,Hamilton_Tegmark_Padmanabhan_2000,%
Percival_et_al_2001,Efstathiou_2dF_team_2002},
and the magnitude-redshift relation of the supernovae Ia
\cite{Hamuy_et_al_1996,Riess_et_al_1998,Perlmutter_et_al_1999}
give strong evidence that the present mean energy density
$\varepsilon_{\hbox{\scriptsize tot}}$ of the universe consists
not only of radiation, baryonic and cold dark matter,
but also of a dominant component with negative pressure
which nowadays is called dark energy.
An obvious candidate for this new energy is Einstein's
cosmological constant $\Lambda$ with a corresponding constant
energy density $\varepsilon_\Lambda=\Lambda c^4/(8\pi G)$ and
negative pressure $p_\Lambda=-\varepsilon_\Lambda$,
assuming a positive cosmological constant.
The associated cosmological models are known as $\Lambda$CDM models.

An alternative explanation for the missing energy is {\it quintessence},
where the dark energy density is identified with the energy density
$\varepsilon_\phi$ (associated with a negative pressure $p_\phi$)
arising from a scalar (quintessence) field $\phi$
(see \cite{Peebles_Ratra_2002} for a recent review).
Quintessence can be considered as a natural generalization of
the cosmological constant to the case of a time-dependent $\Lambda(t)$
with an associated time-dependent pressure.

Many cosmologists seem to accept as established that the universe
is flat corresponding to $k=0$ and
$\Omega_{\hbox{\scriptsize tot}} :=
\varepsilon_{\hbox{\scriptsize tot}} /
\varepsilon_{\hbox{\scriptsize crit}}=1$.
Here we address the question: 
do the recent observations really establish
that our universe is spatially flat?
It is demonstrated that present data are {\it consistent}
with certain quintessence models possessing a constant equation of state
in a hyperbolic universe, i.\,e.\ with negative spatial curvature,
$k=-1$, corresponding to $\Omega_{\hbox{\scriptsize tot}} < 1$.
Our result is the consequence of the important observation that
there exists a {\it degeneracy} in the space of the relevant cosmological
parameters $(w_\phi, \Omega_\phi, \Omega_{\hbox{\scriptsize c}})$
which are introduced below.

Our background model is the standard cosmological model based on a
Friedmann--Lema\^{\i}tre universe with the Robertson-Walker metric
\begin{equation}
\label{Eq:metric}
ds^2 \; = \; a^2(\eta) \, \left( d\eta^2 - \gamma_{ij} dx^i dx^j\right)
\hspace{10pt} ,
\end{equation}
where $\gamma_{ij}$ denotes the spatial hyperbolic metric,
and $a(\eta)$ is the cosmic scale factor as a function of
conformal time $\eta$.
Then the Friedmann equation reads $(a' := da/d\eta, c=1)$
\begin{equation}
\label{Eq:Friedmann}
H^2 := \left(\frac{a'}{a^2}\right)^2 \; = \;
\frac{8\pi G}3 \varepsilon_{\hbox{\scriptsize tot}} + \frac 1{a^2}
\hspace{10pt} ,
\end{equation}
where $H=H(\eta)$ is the Hubble parameter, and the last term in
Eq.~(\ref{Eq:Friedmann}) is the curvature term for $k=-1$.
Furthermore, $\varepsilon_{\hbox{\scriptsize tot}} =
\varepsilon_{\hbox{\scriptsize r}} +
\varepsilon_{\hbox{\scriptsize m}} + \varepsilon_\phi$,
where $\varepsilon_{\hbox{\scriptsize r}}$ denotes the energy density of
``radiation'', i.\,e.\ of the relativistic components according to
photons and three massless neutrinos;
$\varepsilon_{\hbox{\scriptsize m}} =
\varepsilon_{\hbox{\scriptsize b}} +
\varepsilon_{\hbox{\scriptsize cdm}}$
is the energy density of non-relativistic ``matter'' consisting of
baryonic matter, $\varepsilon_{\hbox{\scriptsize b}}$, and
cold dark matter, $\varepsilon_{\hbox{\scriptsize cdm}}$,
and $\varepsilon_\phi$ is the energy density of the dark energy
due to the quintessence field $\phi$.
(In Sect.\,\ref{Potentials_with_Cosmological_Constant}, we will also
include the energy density due to a cosmological constant.)
For our later discussions of the time-dependence
$\varepsilon_{\hbox{\scriptsize x}} = \varepsilon_{\hbox{\scriptsize x}}(\eta)$
of the various energy components
$({\hbox{x}}={\hbox{r}}, {\hbox{m}}, \phi)$, it is important to notice
that the initial conditions to be imposed on the Friedmann equation
(\ref{Eq:Friedmann}) are in the case of negative curvature uniquely
given by $a(0)=0$ and
$a'(0)=\Omega_{\hbox{\scriptsize r}}^{1/2}
(1-\Omega_{\hbox{\scriptsize tot}})^{-1} H_0^{-1}$
with $H_0 = H(\eta_0)$ being the Hubble constant.
Here and in the following, we use the dimensionless density
parameters $\Omega_{\hbox{\scriptsize x}}(\eta) :=
\varepsilon_{\hbox{\scriptsize x}}(\eta) /
\varepsilon_{\hbox{\scriptsize crit}}(\eta)$ with
$\varepsilon_{\hbox{\scriptsize crit}}(\eta) = 3H^2/(8\pi G)$
and $\Omega_{\hbox{\scriptsize x}} := \Omega_{\hbox{\scriptsize x}}(\eta_0)$
denoting their present values.

In quintessence models, the energy density $\varepsilon_\phi(\eta)$
and the pressure $p_\phi(\eta)$ of the dark energy are determined by
the quintessence potential $V(\phi)$
\begin{equation}
\label{Eq:eos_phi}
\varepsilon_\phi \; = \; \frac{1}{2 a^2} \, {\phi'}^2 + V(\phi)
\hspace{10pt} , \hspace{10pt}
p_\phi \; = \; \frac{1}{2 a^2} \, {\phi'}^2 - V(\phi)
\hspace{10pt} ,
\end{equation}
or equivalently by the equation of state
\begin{equation}
\label{Eq:eos_w}
w_\phi(\eta) \; = \; \frac{p_\phi(\eta)}{\varepsilon_\phi(\eta)}
\hspace{10pt} .
\end{equation}
The equation of motion of the real, scalar field $\phi(\eta)$ is
\begin{equation}
\label{Eq:equation_of_motion_phi}
\phi'' \, + \, 2 \frac{a'}{a} \phi' \, + \,
a^2 \frac{\partial V(\phi)}{\partial \phi} \; = \; 0
\hspace{10pt} ,
\end{equation}
where it is assumed that $\phi$ couples to matter only through gravitation.
The various energy densities are constrained by the continuity equation
\begin{equation}
\label{Eq:continuity_equation}
\varepsilon_{\hbox{\scriptsize x}}'(\eta) \, + \,
3 (1+w_{\hbox{\scriptsize x}}(\eta)) \, \frac{a'}a \,
\varepsilon_{\hbox{\scriptsize x}}(\eta) \; = \; 0
\hspace{10pt} ,
\end{equation}
with the constant equation of state
$w_{\hbox{\scriptsize r}}=\frac 13$,
$w_{\hbox{\scriptsize m}}=0$ for $\hbox{x}=\hbox{r},\hbox{m}$
and $w_\phi(\eta)$ for $\hbox{x}=\phi$.
It is worthwhile to remark that the quintessence field $\phi$ may be regarded
as a real physical field, or simply as a device for modeling more general
cosmic fluids with negative pressure.

Obviously, there are two complementary approaches:
\begin{itemize}
\item[a)] Given $V(\phi)$ compute $w_\phi(\eta)$,
\item[b)] Given $w_\phi(\eta)$ compute $V(\phi)$,
\end{itemize}
and then make predictions for or compare with cosmological observations.
Among the various potentials studied in the literature
(see, e.\,g.\ \cite{Peebles_Ratra_2002,Brax_Martin_Riazuelo_2000}
and references therein),
we mention only the {\it inverse power-law potential}
\begin{equation}
\label{Eq:potential_power}
V(\phi) \; = \; \frac{A}{(B\phi)^\gamma}
\hspace{10pt} , \hspace{10pt}
\gamma > 0
\hspace{10pt} ,
\end{equation}
and the {\it exponential potential}
\begin{equation}
\label{Eq:potential_exp}
V(\phi) \; = \; A \, \exp(-B \phi)
\hspace{10pt} , \hspace{10pt}
B > 0
\hspace{10pt} .
\end{equation}
The potentials (\ref{Eq:potential_power}) and (\ref{Eq:potential_exp})
can be {\it derived} \cite{Ratra_Peebles_1988}
(see also Sect.\,\ref{General_Properties}),
if one requires a {\it constant} $w_\phi$ during
a given evolution stage of the universe:
in the radiation-dominated epoch one obtains
the inverse power-law potential (\ref{Eq:potential_power}),
whereas in the quintessence-dominated epoch one requires
$w_\phi = \hbox{const.} \leq -1/3$ and then obtains the
exponential potential (\ref{Eq:potential_exp}).

In the following, we concentrate our attention on approach b), where one
specifies the equation of state $w_\phi$ rather than the potential $V(\phi)$.
In order to generalize the standard cosmological models like
$\Lambda$CDM models with as few as possible additional degrees of freedom,
$w_\phi$ will be chosen as constant (as studied for flat universes in,
e.\,g., \cite{Caldwell_Dave_Steinhardt_1998,Wang_Steinhardt_1998})
This contrasts to approach a), where a whole function,
i.\,e.\ the potential $V(\phi)$, increases the freedom of the model
enormously.
Furthermore, since in the standard cosmological theories
$w_{\hbox{\scriptsize x}}$ is constant for the various energy components,
one can also consider a {\it constant} $w_\phi$ for the dark energy component
as the most ``natural'' generalization.
It turns out that the values $w_\phi = -1/3, -2/3$ play a special role
and constitute a kind of new ``exceptional phases'' in addition to the
standard phases characterized by
$w_{\hbox{\scriptsize x}} = \{1/3, 0, -1\}$  for
$\hbox{x} = \{\hbox{r}, \hbox{m},\Lambda\}$.
Such equations of state occur in models based on topological defects
where $w = -1/3$ corresponds to a network of frustrated cosmic strings
\cite{Vilenkin_1984,Spergel_Pen_1997}
and $w = -2/3$ to domain walls
\cite{Durrer_Kunz_Melchiorri_2002,Gangui_2001}.

One purpose of this paper is to derive the potential $V(\phi)$
for the two special equations of state $w_\phi = -1/3, -2/3$.
The restriction to a constant equation of state arises
from the absence of well motivated dark energy models
being based on fundamental physics.
This restriction should not only be considered as a prejudice
but also as an approximation to time-variable equations of state
when one attempts to describe cosmological observations.
Models with a non-constant equation of state lead to nearly
the same CMB anisotropy as corresponding models with an
effective (constant) equation of state
where $w_\phi(\eta)$ is appropriately $\Omega_\phi$-averaged
\cite{Doran_Lilley_Schwindt_Wetterich_2001,Doran_Lilley_2002}.
As argued in \cite{Doran_Lilley_Schwindt_Wetterich_2001},
the location of the acoustic peaks is mainly determined by
$\Omega_\phi(\eta_0)$, the $\Omega_\phi$-weighted average of
$w_\phi(\eta)$ averaged until the present time,
as well as by the average of $\Omega_\phi(\eta)$ until recombination.
The latter is negligible for negative $w_\phi(\eta)$,
i.\,e.\ for non-tracker field models,
since around $z_{\hbox{\scriptsize sls}}\simeq 1100$
the energy density of the quintessence component is then subdominant
compared to that of the background component.
Thus $\Omega_\phi(\eta_0)$ and the $\Omega_\phi$-weighted average of
$w_\phi(\eta)$ suffice to approximately describe the peak structure.
Since the dark energy dominates only recently,
the $\Omega_\phi$-weighted average leads to an averaging
over the recent history.
Therefore, the anisotropy of the CMB is not well suited to probe
the time-dependence of the equation of state $w_\phi(\eta)$.
The main properties of the anisotropy are then determined
by the angular-diameter distance $d_A$ to the surface of last scattering
(see, e.\,g., \cite{Cornish_2000}).
A further dependence of the CMB anisotropy on the dark energy arises through
the late-time integrated Sachs-Wolfe effect which contributes mostly
to low multipoles in the angular power spectrum.
Because of the cosmic variance and a possible contribution
of gravity waves, it is difficult to extract
information on a time-varying equation of state
\cite{Corasaniti_Bassett_Ungarelli_Copeland_2002}.
On the other hand, in classical cosmological tests
based on the luminosity distance, a dark energy component
contributes appreciably only for redshifts $0.2 \lesssim z \lesssim 2$
\cite{Huterer_Turner_2001}
since for higher redshifts the contribution to the total energy density
is too minute (see also Eq.\,(\ref{Eq:z_m_phi})).
Here the difficulties are caused by the luminosity distance
which depends on $w_\phi(\eta)$ through a multiple-integral
which smears out the information on the time-dependence of $w_\phi(\eta)$
\cite{Maor_Brustein_Steinhardt_2001}.
Thus, with the exception of tracker fields, it is very difficult to
observe a time-dependence of $w_\phi(\eta)$ and therefore
one can restrict the discussion to a constant equation of state.

Quintessence models with a constant equation of state
differ from a cosmological constant with $w_\Lambda=-1$ only in the
value of the constant $w_\phi\neq -1$.
If the observations are too close to $w_\phi\simeq w_\Lambda$,
then quintessence models might be superseded by the long-known
cosmological constant, i.\,e.\ by the vacuum energy.
Assuming {\it flat} universes, the current bound is, e.\,g.,
$w_\phi < -0.85$ at 68\% C.L.\ \cite{Bean_Melchiorri_2002}
or even $w_\phi < -0.93$ at $2\sigma$ \cite{Corasaniti_Copeland_2002}.
The caveat is, as we shall show in Sect.\,\ref{Sect_Observations},
the assumed flatness.
Such a value so close to the cosmological constant has, however,
problems to account for the observed number of giant arcs in galaxy clusters.
For a flat $\Lambda$CDM model with $\Omega_{\hbox{\scriptsize m}}=0.3$,
the number of arcs is one order of magnitude too low.
To obtain the right order, one either needs a low-density open model
with $\Omega_{\hbox{\scriptsize m}}=0.3$
or a flat model where the vacuum energy is replaced by a dark energy
component with $w_\phi \simeq -0.6$
\cite{Bartelmann_Meneghetti_Perrotta_Baccigalupi_Moscardini_2002}.
A further variant, not discussed in 
\cite{Bartelmann_Meneghetti_Perrotta_Baccigalupi_Moscardini_2002},
would be an open model with dark energy which should
also produce enough strong lensing to obtain a large number of arcs.
In addition, too few lensed pairs with wide angular separation are observed
in comparison with the prediction of a flat $\Lambda$CDM-model
\cite{Sarbu_Rusin_Ma_2001}.
These strong lensing observations point towards a universe with
negative curvature.

\section{General Formula for the Quintessence Potential}
\label{Sect_General_Formula}

In the following, general properties of quintessence models
having a constant equation of state $w_\phi$ with $w_\phi\in [-1,-1/3]$
and for a universe with negative curvature, i.\,e.\
$\Omega_{\hbox{\scriptsize tot}} < 1$, will be discussed.
Especially, we discuss the properties of the potential $V(\phi)$
which belongs to a given $w_\phi$.
Without loss of generality, we may assume $\phi(\eta) \geq 0$ with the
initial value $\phi_{\hbox{\scriptsize in}} = \phi(0) = 0$.
It then will turn out that the potential $V(\phi)$ is uniquely determined.

Our starting point are the simple relations
\begin{equation}
\label{Eq:potential_w}
V(\phi) \; = \; \frac 12 \left( \varepsilon_\phi - p_\phi\right) \; = \;
\frac{1-w_\phi}2 \, \varepsilon_\phi
\hspace{10pt} ,
\end{equation}
\begin{equation}
\label{Eq:phi_prime}
{\phi'}^2 \; = \; a^2 \, \left( \varepsilon_\phi + p_\phi\right) \; = \;
(1+w_\phi) \, a^2 \, \varepsilon_\phi
\hspace{10pt} ,
\end{equation}
which directly follow from Eqs.\,(\ref{Eq:eos_phi}) and (\ref{Eq:eos_w}).
Since $w_\phi$ is constant, we obtain the following solution for
$\varepsilon_\phi$ from the continuity equation (\ref{Eq:continuity_equation})
\begin{equation}
\label{Eq:epsilon_phi_redshift}
\varepsilon_\phi(\eta) \; = \; \varepsilon_\phi(\eta_0) \,
\left( \frac{a_0}{a(\eta)}\right)^{3(1+w_\phi)}
\hspace{10pt} ,
\end{equation}
where $a_0 = a(\eta_0) = H_0^{-1} (1-\Omega_{\hbox{\scriptsize tot}})^{-1/2}$
is the scale factor of the present epoch.
Inserting (\ref{Eq:epsilon_phi_redshift}) into (\ref{Eq:potential_w}),
we arrive at the following exact expression for the potential
$V(\phi(\eta))$ parameterized in terms of the scale factor $a(\eta)$
\begin{equation}
\label{Eq:potential_scalefactor}
V(\phi(\eta)) \; = \; \tilde V_0 \,
\left(\frac{a_0}{a(\eta)}\right)^{3(1+w_\phi)}
\hspace{10pt} ,
\end{equation}
where the ``potential strength'' $\tilde V_0 = V(\phi(\eta_0))$ is given by
\begin{equation}
\label{Eq:potential_initial}
\tilde V_0 \; = \; \frac{1-w_\phi}2 \, \Omega_\phi \,
\varepsilon_{\hbox{\scriptsize crit}}
\hspace{10pt} ,
\end{equation}
with $\Omega_\phi = \Omega_\phi(\eta_0) =
\varepsilon_\phi(\eta_0)/\varepsilon_{\hbox{\scriptsize crit}}$
and $\varepsilon_{\hbox{\scriptsize crit}} =
\varepsilon_{\hbox{\scriptsize crit}}(\eta_0) = 3H_0^2/(8\pi G)$.
Although Eq.\,(\ref{Eq:potential_scalefactor}) does not yet express
$V(\phi)$ as a function of $\phi$, several important conclusions
can already be drawn from Eqs.\,(\ref{Eq:potential_scalefactor}) and
(\ref{Eq:potential_initial}):
\begin{itemize}
\item[i)]
$V(\phi)$ is always positive, $V(\phi) \geq 0$,
\item[ii)]
$V(\phi(\eta))$ diverges for $w_\phi=\text{const.} > -1$ like
\begin{equation}
\label{Eq:potential_divergence}
V(\phi(\eta)) \; = \; O\left( \frac 1{\eta^{3(1+w_\phi)}} \right)
\end{equation}
in the radiation-dominated epoch, $\eta\to 0$, since
$a(\eta) = O(\eta)$ in this limit,
\item[iii)]
$V(\phi(\eta))$ decreases monotonically with increasing $\eta$,
since $a(\eta)$ increases monotonically,
\item[iv)]
the order of magnitude of the potential strength $\tilde V_0$,
Eq.\,(\ref{Eq:potential_initial}), is determined by the ``small''
energy density $\varepsilon_{\hbox{\scriptsize crit}}$.
\end{itemize}
The divergent behavior (\ref{Eq:potential_divergence}) for $w_\phi>-1$
implies that $V(\phi(\eta))$ does not possess a Taylor expansion at $\eta=0$,
i.\,e.\ at $\phi_{\hbox{\scriptsize in}}$.
This in turn is tightly connected with the fact that assuming a Taylor
expansion at $\phi=\phi_{\hbox{\scriptsize in}}$ necessarily implies
a time-varying $w_\phi$ with $w_\phi(0)=-1$.
(A proof will be given in Sect.\,\ref{General_Properties}.)

As a side-remark, notice that Eqs.\,(\ref{Eq:potential_scalefactor}) and
(\ref{Eq:potential_initial}) are consistent with the standard cosmological
constant, since $w_\phi=-1$ yields
$V(\phi) = \tilde V_0 = \varepsilon_\phi = \text{const.}$
for all $\eta \geq 0$.

In order to determine $V(\phi)$ as a function of $\phi$,
we must replace $a(\eta)$ in Eq.\,(\ref{Eq:potential_scalefactor})
by $A(\phi) := a(\eta(\phi))$, where $\eta(\phi)$ is the inverse of
$\phi(\eta)$.
The function $A(\phi)$ can by computed as follows.
Combining Eqs.\,(\ref{Eq:phi_prime}) and (\ref{Eq:epsilon_phi_redshift}) yields
\begin{equation}
\label{Eq:phi_prime_eta}
\phi'(\eta) \; = \; \alpha \, [a(\eta)]^\delta
\end{equation}
with $\alpha := [(1+w_\phi) \Omega_\phi a_0^{3(1+w_\phi)} \,
\varepsilon_{\hbox{\scriptsize crit}}]^{1/2}$ and
$\delta := -\frac 12 (1+3w_\phi)$.
Here we have chosen the positive square root of Eq.\,(\ref{Eq:phi_prime}),
since with $\phi(0)=0$ and $\phi(\eta)\geq 0$,
the general properties i)--iii) of the potential lead to
$\phi'(\eta) \geq 0$.
(Actually, Eq.\,(\ref{Eq:phi_prime_eta}) implies
$\phi'(\eta) \to 0$ for $\eta \to 0$ and $-1 < w_\phi < - \frac 13$.)
Furthermore, we have
\begin{equation}
\label{Eq:phi_prime_A}
\phi'(\eta) \; = \; \frac{d\phi(\eta(a))}{da} \, a' \; = \;
\frac{d\phi}{dA} \, \sqrt{f(A)}
\hspace{10pt} ,
\end{equation}
where the function $f(a)$ is defined by rewriting the Friedmann equation
(\ref{Eq:Friedmann}) as ${a'}^2 = f(a)$ with
\begin{eqnarray} \nonumber
f(a) & := &
\frac{8\pi G}3 \, \varepsilon_{\hbox{\scriptsize tot}}(\eta) \, a^4 + a^2
\\ & = &
\label{Eq:f_a}
H_0^2 \, \sum_{\hbox{\scriptsize x}=\hbox{\scriptsize r},
\hbox{\scriptsize m},\phi,\hbox{\scriptsize c}}
\Omega_{\hbox{\scriptsize x}} a_0^{3(1+w_{\hbox{\scriptsize x}})} \,
a^{1-3w_{\hbox{\scriptsize x}}}
\hspace{10pt} .
\end{eqnarray}
In the last expression, we have inserted the solutions of the
continuity equation (\ref{Eq:continuity_equation}) for the various
energy components $\hbox{x} = \{\hbox{r}, \hbox{m},\phi\}$ and,
furthermore, have introduced the dimensionless curvature parameter
$\Omega_{\hbox{\scriptsize c}} := (H_0^2 a_0^2)^{-1} =
1 - \Omega_{\hbox{\scriptsize tot}} > 0$ with the corresponding
equation of state $w_{\hbox{\scriptsize c}} = -\frac 13$.
Combining Eqs.\,(\ref{Eq:phi_prime_eta}) and (\ref{Eq:phi_prime_A}) gives
\begin{equation}
\label{Eq:d_phi}
d\phi \; = \; \alpha \, \frac{A^\delta}{\sqrt{f(A)}} \, dA
\hspace{10pt} ,
\end{equation}
which yields by integration
\begin{equation}
\label{Eq:phi_A}
\phi \; = \; \alpha \, \int_0^{A(\phi)} \frac{A^\delta}{\sqrt{f(A)}} \, dA
\hspace{10pt} .
\end{equation}
Since $A(\phi)=a(\eta(\phi))$ tends to zero in the limit $\eta\to 0$,
the integral relation (\ref{Eq:phi_A}) is consistent with our initial
condition $\phi(\eta)\to 0$ in this limit.
Eq.\,(\ref{Eq:phi_A}) determines the cosmic scale factor $A$ as a function
of the quintessence field $\phi$, i.\,e.\ as the inverse function of
the function defined by the integral in Eq.\,(\ref{Eq:phi_A}).
Inserting the solution $A(\phi)$ of (\ref{Eq:phi_A}) into
Eq.\,(\ref{Eq:potential_scalefactor}), leads to our {\it general formula}
\begin{equation}
\label{Eq:potential_A}
V(\phi) \; = \; \tilde V_0 \, 
\left(\frac{a_0}{A(\phi)}\right)^{3(1+w_\phi)}
\end{equation}
for the quintessence potential.

It is convenient to use the dimensionless variable
$y := \frac A{a_0} = \frac 1{z+1}$ as an integration variable
in (\ref{Eq:phi_A}).
Defining the dimensionless function
\begin{equation}
\label{Def:g}
g(y) \; := \; \sum_{\hbox{\scriptsize x}=\hbox{\scriptsize r},
\hbox{\scriptsize m},\phi,\hbox{\scriptsize c}}
\Omega_{\hbox{\scriptsize x}} \, y^{1-3w_{\hbox{\scriptsize x}}}
\hspace{10pt} ,
\end{equation}
we have $f(A) = f(a_0 y) = H_0^2 a_0^4 g(y)$, and Eq.\,(\ref{Eq:phi_A})
takes the final form
\begin{equation}
\label{Eq:phi_final}
B_0 \, \phi \; = \; \int_0^{A(\phi)/a_0} \frac{y^{-\frac 32(w_\phi+\frac 13)}}
{\sqrt{g(y)}} \, dy
\end{equation}
with
\begin{equation}
\label{Def:B_0}
B_0 \; := \; \frac 1{m_P} \, \sqrt{\frac{8\pi}{3(1+w_\phi) \Omega_\phi}}
\hspace{10pt} .
\end{equation}
(Here $m_P := G^{-1/2}$ denotes the Planck mass, $\hbar=1$.)
Note that the combination $B_0 \phi$ is dimensionless as required by
the right-hand side of (\ref{Eq:phi_final}).

Before we come to a discussion of the general properties of the potential
(\ref{Eq:potential_A}), we have to check whether the
solution (\ref{Eq:phi_final}) for $\phi$ together with the potential
(\ref{Eq:potential_A}) solves the equation of motion
(\ref{Eq:equation_of_motion_phi}), which until now has not been used in
our derivation.
For this purpose, it is useful to express the time-dependence of the
various terms in (\ref{Eq:equation_of_motion_phi}) by $a(\eta)$.
Using $\phi'' = \frac{d\phi'}{da} a' = \frac{d\phi'}{da} \sqrt{f(a)}$
and Eq.\,(\ref{Eq:phi_prime_eta}), we obtain
\begin{equation}
\label{Eq:phi_2prime}
\phi'' \; = \; - \, \frac{\alpha \sqrt{f(a)}}{a^{\frac 32(1+w_\phi)}}
\; \frac{1+3w_\phi}2 
\hspace{10pt} .
\end{equation}
Furthermore, with $\frac{\partial V}{\partial \phi} =
\frac{\partial V}{\partial A} \frac{dA}{d\phi}$ and
Eqs.\,(\ref{Eq:d_phi}) and (\ref{Eq:potential_A}) we derive
\begin{equation}
\label{Eq:V_diff}
a^2 \, \frac{\partial V}{\partial \phi} \; = \; 
- \, \frac{\alpha \sqrt{f(a)}}{a^{\frac 32(1+w_\phi)}} \, \beta
\end{equation}
with $\beta := 3(1+w_\phi) \tilde V_0 a_0^{3(1+w_\phi)}/\alpha^2 =
\frac 32(1-w_\phi)$.
Thus, we obtain
\begin{eqnarray} \nonumber
 \phi'' + 2 \frac{a'}a \phi' + a^2 \frac{\partial V}{\partial \phi} 
\; = \; \hspace{-90pt} & & 
\\ & = & \nonumber
- \, \frac{\alpha \sqrt{f(a)}}{a^{\frac 32(1+w_\phi)}} \,
\frac{1+3 w_\phi}2 \, + \,
2 \frac{\sqrt{f(a)}}a \, \frac{\alpha}{a^{\frac{1+3w_\phi}2}}
\\ & & \nonumber \hspace{120pt} - \,
\frac{\alpha \sqrt{f(a)}}{a^{\frac 32(1+w_\phi)}} \, \beta
\\ & = & \nonumber
- \, \frac{\alpha \sqrt{f(a)}}{a^{\frac 32(1+w_\phi)}} \,
\left[ \frac{1+3w_\phi}2 - 2 + \beta \right] \; \equiv \; 0
\hspace{10pt} ,
\end{eqnarray}
which proves that the equation of motion is satisfied, since the
square bracket in the last expression is identically zero.

\section{General Properties of the Quintessence Potential}
\label{General_Properties}

For arbitrary constant values of $w_\phi \in (-1,-\frac 13)$,
it is not possible to obtain from Eq.\,(\ref{Eq:phi_final})
an explicit analytic expression for $A(\phi)$, and thus no explicit
analytic expression exists in the general case for the potential
(\ref{Eq:potential_A}).
(See, however, the special cases $w_\phi=-\frac 13,-\frac 23$
which will be discussed in Sect.\,\ref{Sect_Analytic_Expressions}.)
Nevertheless, it is possible to compute the potential for all values
of $w_\phi$ numerically, and therefore one can calculate various
quantities, which then can be compared with the cosmological observations.
This will be done in Sect.\,\ref{Sect_Observations}.
In this Section, we discuss some general properties which can be deduced
from the formulae derived in Sect.\,\ref{Sect_General_Formula}.

In the radiation-dominated epoch, $\eta\to 0$ or $\phi\to 0$,
we have $A(\phi) \to 0$, and thus Eq.\,(\ref{Eq:phi_final}) gives
($\delta := -\frac 32 (\frac 13+w_\phi) \geq 0$)
\begin{eqnarray} \nonumber
B_0 \, \phi & = & \int_0^{A(\phi)/a_0} \frac{y^\delta}
{\sqrt{\Omega_{\hbox{\scriptsize r}} + \Omega_{\hbox{\scriptsize m}} y
+ O(y^2)}} \, dy
\\ & = & \label{Eq:phi_sing}
\frac 1{\sqrt{\Omega_{\hbox{\scriptsize r}}}(\delta+1)} \,
\left( \frac{A(\phi)}{a_0} \right)^{\delta+1} \, + \, \dots
\hspace{10pt} ,
\end{eqnarray}
which implies
\begin{equation}
\label{Eq:A_sing}
A(\phi) \; = \; a_0 \left[ (\delta+1) B_0 \sqrt{\Omega_{\hbox{\scriptsize r}}}
\; \right]^{\frac 1{\delta+1}} \, \phi^{\frac 1{\delta+1}}
\, + \, \dots
\hspace{3pt} , \hspace{3pt}
\phi \to 0
\hspace{3pt} .
\end{equation}
Inserting this in (\ref{Eq:potential_A}),
gives for the quintessence potential
\begin{equation}
\label{Eq:Pot_limit_0}
V(\phi) \; = \; O\left(\frac 1{\phi^\gamma}\right)
\hspace{10pt} , \hspace{10pt}
\phi \to 0
\hspace{10pt} , \hspace{10pt}
\gamma \; = \; 6 \, \frac{1+w_\phi}{1-3 w_\phi}
\hspace{7pt} .
\end{equation}
For the two exceptional cases $w_\phi = -1/3$ and $-2/3$,
one obtains from (\ref{Eq:Pot_limit_0}) $\gamma=2$ and $2/3$, respectively.
(If radiation is neglected, $\Omega_{\hbox{\scriptsize r}}=0$,
while keeping $\Omega_{\hbox{\scriptsize m}} \neq 0$,
as it is often assumed, one instead obtains (see Eq.\,(\ref{Eq:phi_sing}))
$\gamma = -2(1+w_\phi)/w_\phi$ which yields in the case
$w_\phi=-1/3$ the value $\gamma=4$ instead of $\gamma=2$.)
Thus, the quintessence potential behaves for $\phi\to 0$ as
the inverse power-law potential (\ref{Eq:potential_power}),
where the power $\gamma$ is uniquely given by Eq.\,(\ref{Eq:Pot_limit_0}).

In the quintessence-dominated epoch, one has $a(\eta)\to\infty$
for $\eta\to\eta_\infty$ with $\eta_\infty<\infty$ for
$w_\phi<-\frac 13$ and $\eta_\infty=\infty$ for $w_\phi=-\frac 13$.
This follows immediately from the Friedmann equation
$a'=da/d\eta=\sqrt{f(a)}$, which can be integrated to give
\begin{equation}
\label{Eq:eta_a}
\eta \; = \; \int_0^{a(\eta)} \frac{da}{\sqrt{f(a)}} \; = \;
\sqrt{\Omega_{\hbox{\scriptsize c}}} \int_0^{a(\eta)/a_0}
\frac{dy}{\sqrt{g(y)}}
\hspace{10pt} .
\end{equation}
For $w_\phi<-\frac 13$ this yields
\begin{equation}
\label{Eq:eta_infty}
\eta_\infty =
\sqrt{\Omega_{\hbox{\scriptsize c}}} \int_0^\infty
\frac{dy}{\sqrt{\Omega_\phi y^{1-3w_\phi} + \Omega_{\hbox{\scriptsize c}} y^2
+ \Omega_{\hbox{\scriptsize m}} y + \Omega_{\hbox{\scriptsize r}} }}
< \infty \, .
\end{equation}
Similarly, we obtain from (\ref{Eq:phi_final})
$\phi \to \infty$ in the limit $\eta\to\eta_\infty$ for
$w_\phi \leq - \frac 13$.
To get the leading asymptotic behavior of $A(\phi)$ in this limit,
we have to distinguish again between two cases.
For $w_\phi=-\frac 13$, Eq.\,(\ref{Eq:phi_final}) yields
\begin{eqnarray} \nonumber
B_0 \, \phi & = & \int_0^{A(\phi)/a_0} \frac{dy}
{\sqrt{(\Omega_\phi+\Omega_{\hbox{\scriptsize c}}) y^2 +
\Omega_{\hbox{\scriptsize m}} y + \Omega_{\hbox{\scriptsize r}}}}
\\ & = & \label{Eq:phi_limit_infty_13}
\frac 1{\sqrt{\Omega_\phi+\Omega_{\hbox{\scriptsize c}}}} \,
\ln \left( \frac{A(\phi)}{a_0} \right) \, + \, \dots
\hspace{10pt} ,
\end{eqnarray}
whereas for $-1 < w_\phi < -\frac 13$ on gets
\begin{eqnarray} \nonumber
B_0 \, \phi & = & \int_0^{A(\phi)/a_0} \frac{y^{-\frac 32(w_\phi+\frac 13)}}
{\sqrt{\Omega_\phi y^{1-3w_\phi} + O(y^2)}} \, dy
\\ & = & \label{Eq:phi_limit_infty}
\frac 1{\sqrt{\Omega_\phi}} \,
\ln \left( \frac{A(\phi)}{a_0} \right) \, + \, \dots
\hspace{10pt} .
\end{eqnarray}
Thus, we obtain from Eqs.\,(\ref{Eq:phi_limit_infty_13}),
(\ref{Eq:phi_limit_infty}) and (\ref{Def:B_0})
\begin{equation}
\label{Eq:A_infty}
A(\phi) \; = \; O(e^{B\phi})
\hspace{10pt} , \hspace{10pt}
\phi \to \infty
\hspace{10pt} ,
\end{equation}
with
\begin{equation}
\label{Def:B}
B := \left\{ \begin{matrix}
& \hspace*{-5pt}B_0 & \hspace*{-6pt} \sqrt{\Omega_\phi+\Omega_{\hbox{\scriptsize c}}} \, = \,
\frac{2\sqrt\pi}{m_P} \,
\sqrt{1+\frac{\Omega_{\hbox{\scriptsize c}}}{\Omega_\phi}}
\hbox{ for } w_\phi = -\frac 13 \\
& \hspace*{-5pt}B_0 & \hspace*{-6pt} \sqrt{\Omega_\phi} \, = \,
\frac{2\sqrt\pi}{m_P} \, \sqrt{\frac 2{3(1+w_\phi)}}
\hbox{ for } -1 < w_\phi < -\frac 13
\end{matrix} \right.
\hspace{3pt} .
\end{equation}
Inserting the asymptotic behavior (\ref{Eq:A_infty}) in (\ref{Eq:potential_A}),
gives
\begin{equation}
\label{Eq:V_limit_infty}
V(\phi) \; = \; O\left(e^{-\mu \phi}\right)
\hspace{10pt} , \hspace{10pt}
\phi \to \infty
\hspace{10pt} ,
\end{equation}
with
\begin{eqnarray} \nonumber
\mu & := & 3(1+w_\phi) B
\\ & = & \label{Def:mu}
\frac{2\sqrt\pi}{m_P}
\left\{ \begin{matrix}
2\sqrt{1+\frac{\Omega_{\hbox{\scriptsize c}}}{\Omega_\phi}}
\hbox{ for } w_\phi = -\frac 13 \\
 \sqrt{6(1+w_\phi)}
\hbox{ for } -1 < w_\phi < -\frac 13
\end{matrix} \right. .
\end{eqnarray}
Thus, the quintessence potential behaves for $\phi\to\infty$ as
the exponential potential (\ref{Eq:potential_exp}),
where the exponent $\mu$ is uniquely given by (\ref{Def:mu}).

We conclude that for $w_\phi=\hbox{const.}\leq - 1/3$ and
$\Omega_{\hbox{\scriptsize tot}} = 1 - \Omega_{\hbox{\scriptsize c}} < 1$
the derived quintessence potential (\ref{Eq:potential_A}) {\it interpolates}
between the inverse power-law potential (\ref{Eq:potential_power})
for $\phi\to 0$ and the exponential potential (\ref{Eq:potential_exp})
for $\phi\to \infty$.
There arises then the question: does there exist a closed analytic expression
for $V(\phi)$ valid for all $\phi\geq 0$?
In Sect.\,\ref{Sect_Analytic_Expressions} we shall study the special
cases $w_\phi=-\frac 13,-\frac 23$
and shall show that for these cases there exist, indeed, exact analytic
expressions for the potential.

Before closing this section, we would like to show
that the divergent behavior (\ref{Eq:Pot_limit_0}) for $\phi\to 0$
(see also Eq.\,(\ref{Eq:potential_divergence})), which makes it impossible
to expand $V(\phi)$ at $\phi=0$ into a Taylor series,
is a consequence of the assumption that the quintessence field has a
constant equation of state.

Let us suppose, on the contrary, that we have a potential $V(\phi)$
which possesses a well-defined Taylor expansion at some initial field
$\phi_{\hbox{\scriptsize in}}=\phi(0)$, i.\,e.\
\begin{equation}
\label{Eq:V_Taylor}
V(\phi) \; = \; v_0 + v_1(\phi-\phi_{\hbox{\scriptsize in}}) \, + \,
O\left( (\phi-\phi_{\hbox{\scriptsize in}})^2\right)
\end{equation}
with $v_0:=V(\phi_{\hbox{\scriptsize in}})>0$,
$v_1:=\frac{\partial V}{\partial \phi}(\phi_{\hbox{\scriptsize in}})\neq 0$,
while $\phi(\eta)$ has at $\eta=0$ the expansion $(b\neq 0, \sigma>0)$
\begin{equation}
\label{Eq:phi_in_case_Taylor}
\phi(\eta) \; = \; \phi_{\hbox{\scriptsize in}} \, + \,
b \, \eta^\sigma(1+O(\eta))
\hspace{10pt} .
\end{equation}
Using the expansions (\ref{Eq:V_Taylor}), (\ref{Eq:phi_in_case_Taylor})
and the well-known expansion of the scale factor in the radiation-dominated
epoch, $a(\eta)=a'(0)\, \eta + O(\eta^2)$, it is not difficult to see
that the equation of motion (\ref{Eq:equation_of_motion_phi}) is satisfied
iff $\sigma=4$ and $b=-\frac{v_1}{20} {a'}^2(0)$.

Now, let us consider the equation of state (\ref{Eq:eos_w}), which by means
of (\ref{Eq:eos_phi}) can be rewritten as
\begin{equation}
\label{Eq:w_r}
w_\phi(\eta) \; = \; - \, \frac{1-r_\phi(\eta)}{1+r_\phi(\eta)}
\hspace{10pt} , \hspace{10pt}
r_\phi(\eta) \; := \; \frac{T(\eta)}{V(\phi(\eta))}
\hspace{10pt} .
\end{equation}
Here $r_\phi$ denotes the ratio of the kinetic energy
$T(\eta):={\phi'}^2/(2a^2)$ of the quintessence field to its potential
energy $V$.
From the above equations, one derives that $T$ vanishes like $O(\eta^4)$,
which leads to
\begin{equation}
\label{Eq:r_phi_Taylor}
r_\phi(\eta) \; = \;
\frac{1}{50} \, \frac{v_1^2}{v_0} \, {a'}^2(0) \, \eta^4 \, + \, O(\eta^5)
\end{equation}
and
\begin{equation}
\label{Eq:w_phi_Taylor}
w_\phi(\eta) \; = \; -1 \, + \,
\frac{1}{25} \, \frac{v_1^2}{v_0} \, {a'}^2(0) \, \eta^4 \, + \, O(\eta^5)
\end{equation}
in the limit $\eta\to 0$.
Thus, in such a model the quintessence component is strongly suppressed at
early times and practically indistinguishable from a cosmological constant
with $w_\Lambda=-1$.
However, with increasing time $\eta$, $w_\phi(\eta)$ increases and therefore
cannot stay constant at a value $w_\phi \in [-1,-\frac 13]$ for all times.

An example of a quintessence model of this type is given by the
exponential potential (\ref{Eq:potential_exp}) for which
$\phi_{\hbox{\scriptsize in}}$ can be set to zero without loss of generality.
This model has been studied in detail, e.\,g., in \cite{Aurich_Steiner_2002a}
for hyperbolic universes.
Here $w_\phi(\eta)$ starts out at $-1$ and then
approaches zero in the flat case or $-\frac 13$ in the hyperbolic case.
This shows clearly, as stated already, why our assumption of a constant
equation of state necessarily implies an inverse power-law divergence
of $V(\phi)$ in the radiation-dominated epoch,
see Eq.\,(\ref{Eq:Pot_limit_0}).

\section{Time-Evolution of the Energy Density Parameters
$\Omega_{\hbox{\scriptsize x}}(\eta)$}
\label{Sect_Energy_Density}

In this Section, we study the time-dependence of the dimensionless energy
density parameters $\Omega_{\hbox{\scriptsize x}}(\eta) =
\varepsilon_{\hbox{\scriptsize x}}(\eta)/
\varepsilon_{\hbox{\scriptsize crit}}(\eta)$
for the various energy components $\hbox{x} = \{\hbox{r}, \hbox{m},\phi\}$
as well as for $\hbox{x} = \hbox{tot}$ referring to the
total energy density.
From the continuity equation (\ref{Eq:continuity_equation})
we obtain for $\hbox{x} = \{\hbox{r}, \hbox{m},\phi\}$
\begin{equation}
\label{Eq:epsilon_redshift}
\varepsilon_{\hbox{\scriptsize x}}(\eta) \; = \;
\varepsilon_{\hbox{\scriptsize x}}(\eta_0) \,
\left(\frac{a_0}{a(\eta)}\right)^{3(1+w_{\hbox{\scriptsize x}})}
\hspace{10pt} ,
\end{equation}
since all three components are assumed to possess a constant equation
of state $w_{\hbox{\scriptsize x}}$.
Expressing the Hubble parameter $H(\eta)$ entering
$\varepsilon_{\hbox{\scriptsize crit}}(\eta)$ by the Friedmann equation
(\ref{Eq:Friedmann}) and using the variable
$y := \frac a{a_0} = \frac 1{z+1}$ together with the function $g(y)$,
see Eq.\,(\ref{Def:g}), one obtains
\begin{equation}
\label{Eq:Omega_redshift_g}
\Omega_{\hbox{\scriptsize x}}(\eta) \; = \;
\frac{y^{1-3w_{\hbox{\scriptsize x}}}}{g(y)} \, \Omega_{\hbox{\scriptsize x}}
\hspace{10pt} , \hspace{10pt}
\hbox{x} = \{\hbox{r}, \hbox{m},\phi\}
\hspace{10pt} .
\end{equation}
The general relation (\ref{Eq:Omega_redshift_g}) determines the
time-evolution of the parameters $\Omega_{\hbox{\scriptsize x}}(\eta)$
in terms of their present values
$\Omega_{\hbox{\scriptsize x}} = \Omega_{\hbox{\scriptsize x}}(\eta_0)$
parameterized by the scale factor $a(\eta)$, respectively
the redshift $z$.
It follows immediately $\Omega_{\hbox{\scriptsize r}}(0)=1$ and
$\Omega_{\hbox{\scriptsize m}}(0)=\Omega_\phi(0)=0$.
With $\Omega_{\hbox{\scriptsize tot}}(\eta) =
\sum_{\hbox{\scriptsize x}=\hbox{\scriptsize r},\hbox{\scriptsize m},\phi}
\Omega_{\hbox{\scriptsize x}}(\eta)$
and Eqs.\,(\ref{Eq:Omega_redshift_g}) and (\ref{Def:g}) one derives
\begin{equation}
\label{Eq:Omega_tot_limit_0}
\Omega_{\hbox{\scriptsize tot}}(\eta) \; = \;
1 \, - \, \frac{y^2}{g(y)} \, (1-\Omega_{\hbox{\scriptsize tot}})
\end{equation}
satisfying $\Omega_{\hbox{\scriptsize tot}}(0)=1$.
(Notice that
$g(1)=\Omega_{\hbox{\scriptsize tot}}+\Omega_{\hbox{\scriptsize c}}=1$.)

To analyze the behavior of Eqs.\,(\ref{Eq:Omega_redshift_g}) and
(\ref{Eq:Omega_tot_limit_0}) in the radiation-dominated epoch,
we observe that Eq.\,(\ref{Eq:eta_a}) determines in the case
$w_\phi \leq - \frac 13$ the first two terms of the scale factor
$a(\eta)$ uniquely
\begin{equation}
\label{Eq:y_Taylor}
y \; = \; \frac{a(\eta)}{a_0} \; = \;
\frac{a'(0)}{a_0}\left( \eta + \frac{\eta^2}{2\hat\eta} + O(\eta^3)\right)
\end{equation}
with $\hat \eta :=
2\sqrt{\Omega_{\hbox{\scriptsize r}}\Omega_{\hbox{\scriptsize c}}}/
\Omega_{\hbox{\scriptsize m}} \simeq (1+\sqrt 2) \eta_{\hbox{\scriptsize eq}}$,
where $\eta_{\hbox{\scriptsize eq}}$ denotes the conformal time at
matter-radiation equality.
We then obtain for $\eta\to 0$
\begin{equation}
\label{Eq:Omega_eta_small}
\begin{matrix}
\Omega_{\hbox{\scriptsize r}}(\eta) & = & 1 - 2(\eta/\widehat{\eta})
+ O(\eta^2) \\
\Omega_{\hbox{\scriptsize m}}(\eta) & = & 2(\eta/\widehat{\eta})
+ O(\eta^2) \\
\Omega_\phi(\eta) & = & O\left(\eta^{1-3 w_\phi}\right) \\
\Omega_{\hbox{\scriptsize tot}}(\eta) & = &
1 - \eta^2 + O(\eta^{1-3 w_\phi})
\end{matrix}
\hspace{10pt} .
\end{equation}
The crucial point to observe is that the quintessence component
$\Omega_\phi(\eta)$ is at early times suppressed
(the more the smaller $w_\phi$ is) such that
$\Omega_\phi(\eta_{\hbox{\scriptsize BBN}}) \ll 10^{-5}$
and thus does not interfere
with the strong constraints coming from the big-bang nucleosynthesis (BBN).

\begin{figure}
\begin{center}
\vspace*{-15pt}
\begin{minipage}{8cm}
\includegraphics[width=8.0cm,height=5.0cm]{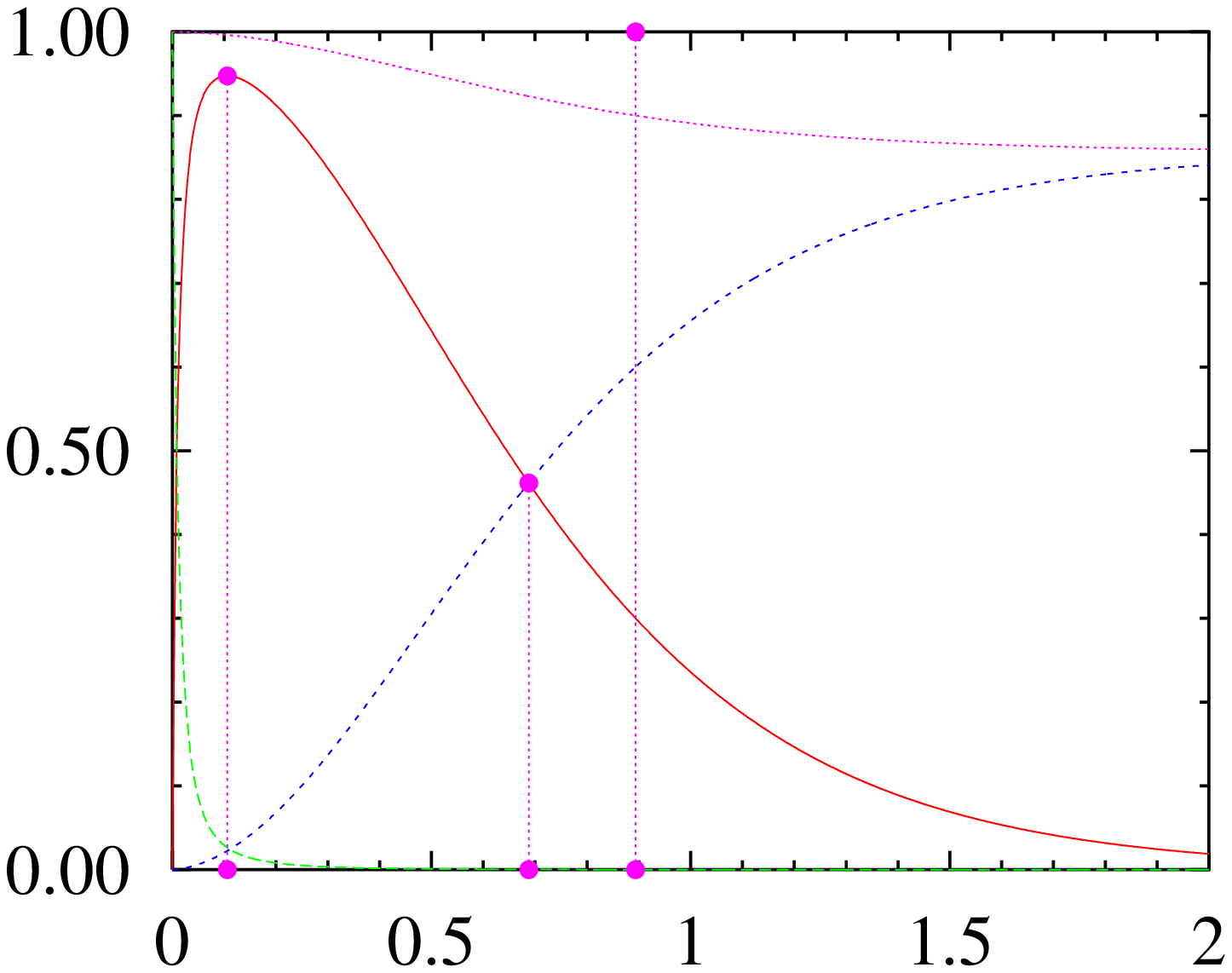}
\put(-30,7){$\eta$}
\put(-205,114){$\Omega_{\hbox{\scriptsize x}}(\eta)$}
\put(-104,70){$\eta_0$}
\put(-140,32){$\eta_{\hbox{\scriptsize m}\phi}$}
\put(-167,70){$\eta_{\hbox{\scriptsize m}}$}
\put(-173,30){r}
\put(-60,30){m}
\put(-50,100){$\phi$}
\put(-50,118){tot}
\end{minipage}
\vspace*{-20pt}
\end{center}
\caption{\label{Fig:Omega_eta}
The energy densities $\Omega_{\hbox{\scriptsize x}}(\eta)$ are shown for
the quintessence potential (\ref{Eq:V_13}) for
$w_\phi = -\frac 13$ using the parameters given in
Sect.\,\ref{Sect_Observations}.
}
\end{figure}

At late times, $\eta\to\eta_\infty$, respectively $z\to -1$,
we obtain in the case $w_\phi = - \frac 13$
\begin{equation}
\label{Eq:Omega_eta_large_13}
\begin{matrix}
\Omega_{\hbox{\scriptsize r}}(\eta) & = &
\frac{\Omega_{\hbox{\scriptsize r}}}{\Omega_{\hbox{\scriptsize c}}+\Omega_\phi}
\, (1+z)^2 \, + \, \dots \; \to \; 0 \hspace{41pt} \\
\Omega_{\hbox{\scriptsize m}}(\eta) & = & 
\frac{\Omega_{\hbox{\scriptsize m}}}{\Omega_{\hbox{\scriptsize c}}+\Omega_\phi}
\, (1+z) \, + \, \dots \; \to \; 0 \hspace{48pt} \\
\Omega_\phi(\eta) & \simeq & \Omega_{\hbox{\scriptsize tot}}(\eta) \; = \;
1 - \frac{\Omega_{\hbox{\scriptsize c}}}{\Omega_{\hbox{\scriptsize c}}
+\Omega_\phi} \, + \, \dots < 1 \hspace{18pt}
\end{matrix} ,
\end{equation}
whereas for $w_\phi < - \frac 13$ one derives
\begin{equation}
\label{Eq:Omega_eta_large}
\begin{matrix}
\Omega_{\hbox{\scriptsize r}}(\eta) & = &
\frac{\Omega_{\hbox{\scriptsize r}}}{\Omega_\phi} \, (1+z)^{1-3 w_\phi}
\, + \, \dots \; \to \; 0 \hspace{56pt} \\
\Omega_{\hbox{\scriptsize m}}(\eta) & = & 
\frac{\Omega_{\hbox{\scriptsize m}}}{\Omega_\phi} \, (1+z)^{-3 w_\phi }
\, + \, \dots \; \to \; 0 \hspace{63pt} \\
\Omega_\phi(\eta) & \simeq & \Omega_{\hbox{\scriptsize tot}}(\eta) \, = \,
1 - \frac{\Omega_{\hbox{\scriptsize c}}}{\Omega_\phi} \, (1+z)^{-(1+3 w_\phi)}
+ \dots \to 1
\end{matrix} .
\end{equation}
We see that in all models the quintessence component dominates at low redshift
(see Fig.\ref{Fig:Omega_eta} in the case $w_\phi=-\frac 13$).

In the case $w_\phi<-\frac13$, the universe becomes asymptotically flat,
whereas for $w_\phi=-\frac13$ it stays forever hyperbolic with
$\Omega_{\hbox{\scriptsize tot}}(\infty)<1$,
see Eq.\,(\ref{Eq:Omega_eta_large_13}).
Radiation-quintessence equality occurs at $z=z_{{\hbox{\scriptsize r}}\phi}$
with
\begin{equation}
\label{Eq:equality_phi_r}
z_{{\hbox{\scriptsize r}}\phi} \; = \;
\left( \frac{\Omega_\phi}{\Omega_{\hbox{\scriptsize r}}}
\right)^{\frac 1{1-3w_\phi}} \, - \, 1
\hspace{10pt} .
\end{equation}
The matter component takes its maximum value at a redshift value
$z_{\hbox{\scriptsize m}}$, which can be obtained as a solution of the
equation
\begin{equation}
\label{Eq:z_m}
\Omega_{\hbox{\scriptsize r}} (1+z_{\hbox{\scriptsize m}})^{1-3w_\phi}
\, - \,
\Omega_{\hbox{\scriptsize c}} (1+z_{\hbox{\scriptsize m}})^{-(1+3w_\phi)}
\; = \;
-3 \, w_\phi \, \Omega_\phi
\hspace{10pt} .
\end{equation}
Matter-quintessence equality holds at $z=z_{\hbox{\scriptsize m}\phi}$ with
\begin{equation}
\label{Eq:z_m_phi}
z_{\hbox{\scriptsize m}\phi} \; = \; \left(\frac{\Omega_\phi}
{\Omega_{\hbox{\scriptsize m}}}\right)^{-1/(3w_\phi)} \, - \, 1
\hspace{10pt} .
\end{equation}
Fig.\ref{Fig:Omega_eta} shows the time-evolution of the energy densities
$\Omega_{\hbox{\scriptsize x}}(\eta)$ for $w_\phi=-\frac 13$
using the parameters discussed in Sect.\,\ref{Sect_Observations}.
During the epoch of quintessence dominance the growth of
linear perturbations is suppressed.
This suppression increases thus with increasing $w_\phi$
since then dark energy dominates earlier.
In cases of too early dark energy dominance, one needs either a
large bias parameter or models with a large CDM-contribution
in order to obtain a power spectrum $P(k)$ of large scale structure
that is consistent with the observations.

\section{Explicit Analytic Expressions for the Quintessence Potential}
\label{Sect_Analytic_Expressions}

Formula (\ref{Eq:potential_A}) gives the quintessence potential $V(\phi)$
in terms of the scale factor $A(\phi)$, which in turn has to be obtained
by ``inversion'' of the integral appearing in Eq.\,(\ref{Eq:phi_final})
\begin{eqnarray}
\label{Def:I_w}
I_w(u) & = & I_w(u;\Omega_{\hbox{\scriptsize r}},\Omega_{\hbox{\scriptsize m}},
\Omega_{\hbox{\scriptsize c}},\Omega_\phi)
\\ & := &  \nonumber
\int_0^u \frac{y^{-\frac 12 - \frac 32 w}}
{\sqrt{\Omega_\phi y^{1-3w} + \Omega_{\hbox{\scriptsize c}} y^2 +
\Omega_{\hbox{\scriptsize m}} y + \Omega_{\hbox{\scriptsize r}}}} \; dy
\hspace{10pt} ,
\end{eqnarray}
where in the following we assume $u \geq 0$, $w\in[-1,-\frac 13]$ and
$\Omega_{\hbox{\scriptsize x}}>0$ for
$\hbox{x} = \hbox{r}, \hbox{m}, \hbox{c},\phi$.
The problem of integration  posed by (\ref{Def:I_w}) is in general
insoluble by means of elementary functions.
For special values of $w$, however, it is possible to give explicit
solutions to our problem.

\subsection{The Potential for $w_\phi=-\frac 13$}
\label{Sect_Analytic_Expressions_13}

In the case of a quintessence component belonging to a constant
equation of state, $w_\phi=-\frac 13$, the integral (\ref{Def:I_w})
is particularly simple
$(\tilde \Omega_\phi := \Omega_\phi + \Omega_{\hbox{\scriptsize c}})$
\begin{equation}
\label{Eq:I_13}
I_{-\frac 13}\left(\frac{A(\phi)}{a_0}\right) \; = \;
\frac 1{\sqrt{\tilde \Omega_\phi}} \, \ln\left(\frac XY \right)
\hspace{10pt} ,
\end{equation}
where we have introduced the abbreviations
\begin{eqnarray} \nonumber
X & := &  2 \sqrt{Z} \, + \, 2 \tilde \Omega_\phi \, \left(\frac{A}{a_0}\right)
\, + \, \Omega_{\hbox{\scriptsize m}} 
\\ Y & := & \label{Def:XYZ}
2 \sqrt{\tilde \Omega_\phi\Omega_{\hbox{\scriptsize r}}} \, + \,
\Omega_{\hbox{\scriptsize m}}
\\ Z & := & \nonumber
\tilde \Omega_\phi \left( \Omega_{\hbox{\scriptsize r}} \, + \,
\Omega_{\hbox{\scriptsize m}}\, \left(\frac{A}{a_0}\right) \, + \,
\tilde \Omega_\phi\, \left(\frac{A}{a_0}\right)^2 \right)
\hspace{10pt} .
\end{eqnarray}
From Eqs.\,(\ref{Eq:I_13}) and (\ref{Eq:phi_final}) one gets
\begin{equation}
\label{Eq:X_13}
X \; = \; Y \, e^{B\phi}
\hspace{10pt} ,
\end{equation}
where $B$ is given in Eq.\,(\ref{Def:B}) for $w_\phi=-\frac 13$.
Furthermore, we obtain from (\ref{Def:XYZ})
$(X-\Omega_{\hbox{\scriptsize m}}) -
2\tilde \Omega_\phi\, \left(\frac{A}{a_0}\right) = 2 \sqrt Z$,
and squaring this equation yields
\begin{eqnarray} \nonumber
(X-\Omega_{\hbox{\scriptsize m}})^2 \, - \,
4\tilde \Omega_\phi\, (X-\Omega_{\hbox{\scriptsize m}})\,
\left(\frac{A}{a_0}\right) \, + \,
4\tilde \Omega_\phi^2 \left(\frac{A}{a_0}\right)^2
& = & \\
& & \nonumber \hspace{-180pt}
4 \tilde \Omega_\phi \left( \Omega_{\hbox{\scriptsize r}} \, + \,
\Omega_{\hbox{\scriptsize m}}\, \left(\frac{A}{a_0}\right) \, + \,
\tilde \Omega_\phi\, \left(\frac{A}{a_0}\right)^2 \right)
\hspace{10pt} ,
\end{eqnarray}
which can easily be solved for $A/a_0$ giving
$$
\frac{A(\phi)}{a_0} \; = \; 
\frac 1{4\tilde \Omega_\phi} \left( X - 2\Omega_{\hbox{\scriptsize m}} +
(\Omega_{\hbox{\scriptsize m}}^2 -
4 \tilde \Omega_\phi\Omega_{\hbox{\scriptsize r}}) \, X^{-1} \right)
\hspace{10pt} .
$$
Replacing then $X$ by the solution (\ref{Eq:X_13}),
leads to the following explicit expression for the scale factor $A(\phi)$
as a function of the quintessence field $\phi$
\begin{equation}
\label{Eq:A_13}
\frac{A(\phi)}{a_0} \; = \; \frac{\Omega_{\hbox{\scriptsize m}}}
{2(\Omega_\phi+\Omega_{\hbox{\scriptsize c}})} \,
\left( \tilde \eta \sinh(B\phi) + \cosh(B\phi) - 1 \right)
\end{equation}
with $\tilde \eta := \frac 2{\Omega_{\hbox{\scriptsize m}}}
\sqrt{\Omega_{\hbox{\scriptsize r}}
(\Omega_\phi+\Omega_{\hbox{\scriptsize c}})}$.
It remains to insert the result (\ref{Eq:A_13}) into the general
formula (\ref{Eq:potential_A}), to arrive at the
{\it exact quintessence potential} for $w_\phi=-\frac 13$
\begin{equation}
\label{Eq:V_13}
V(\phi) \; = \; \frac{V_0}
{\left[ \tilde \eta \sinh(B\phi) + \cosh(B\phi) - 1 \right]^2}
\hspace{10pt} ,
\end{equation}
where we defined the potential strength
$V_0 = \frac 83 \varepsilon_{\hbox{\scriptsize crit}} \Omega_\phi
(\Omega_\phi+\Omega_{\hbox{\scriptsize c}})^2
/\Omega_{\hbox{\scriptsize m}}^2$.
Thus, in the case $w_\phi=-\frac 13$ the quintessence potential
is completely known, not only in its functional form (\ref{Eq:V_13}),
but also with its para\-meters which are uniquely determined by $H_0$
and the present cosmological density parameters
$\Omega_{\hbox{\scriptsize r}}$, $\Omega_{\hbox{\scriptsize m}}$,
$\Omega_\phi$ and $\Omega_{\hbox{\scriptsize c}} = 1 -
\Omega_{\hbox{\scriptsize r}} - \Omega_{\hbox{\scriptsize m}} - \Omega_\phi$.
The potential (\ref{Eq:V_13}) and therefore the cosmic evolution is
governed by {\it two very different energy scales}:
the small critical energy $\varepsilon_{\hbox{\scriptsize crit}}$ and
the huge Planck mass $m_P$!
Explicitly, one derives from (\ref{Eq:V_13})
\begin{equation}
\label{Eq:V_13_limit}
V(\phi) \; = \; \left\{ \begin{matrix}
\frac{V_0}{(\tilde \eta B)^2} \, \frac 1{\phi^2}
\hbox{ for } \phi \to 0 \\
\frac{4V_0}{(1+\tilde \eta)^2} \, e^{-2B\phi}
\hbox{ for } \phi \to \infty
\end{matrix} \right.
\hspace{10pt} ,
\end{equation}
which is in accordance with the general behavior given by
Eqs.\,(\ref{Eq:Pot_limit_0}) and (\ref{Eq:V_limit_infty}), respectively.

It is interesting to consider the limit of the potential (\ref{Eq:V_13})
for vanishing radiation.
In this limit, we obtain the following {\it exact quintessence potential}
for a two-component model consisting of matter and quintessence
with $w_\phi =-\frac 13$
\begin{equation}
\label{Eq:V_13_without_radiation}
V(\phi) \; = \; \frac{V_0}4 \, \frac 1{\sinh^4\left(\frac{B\phi}2\right)}
\hspace{10pt} .
\end{equation}
In Sect.\,\ref{V_23_Or_0}, we shall show that this potential leads to
an {\it almost} constant equation of state $w_\phi(\eta)$ with
$w_\phi(0)=-\frac 19$ and $w_\phi(\eta)\to -\frac 13$ for $\eta\to\infty$,
if (\ref{Eq:V_13_without_radiation}) is assumed to govern the
time-evolution of a quintessence field, where a full three-component
background model consisting of radiation, matter and quintessence is employed.
(See Sect.\,\ref{V_23_Or_0} for details.)

\subsection{The Potential for $w_\phi=-\frac 23$ and
$\Omega_{\hbox{\scriptsize r}}\equiv 0$}
\label{V_23_Or_0}

If the quintessence component possesses a constant equation of state,
$w_\phi=-\frac 23$, the relevant integral (\ref{Def:I_w}) reads
\begin{equation}
\label{Eq:I_23}
I_{-\frac 23}(u) \; = \;
\int_0^u \frac{\sqrt{y}}
{\sqrt{\Omega_\phi y^3 + \Omega_{\hbox{\scriptsize c}} y^2 +
\Omega_{\hbox{\scriptsize m}} y + \Omega_{\hbox{\scriptsize r}}}} \; dy
\hspace{10pt} .
\end{equation}
The integral (\ref{Eq:I_23}) can be transformed into an elliptic integral
of the third kind which can be expressed in terms of the Weierstrass
$\sigma$- and $\zeta$-function.
There exists, however, no simple inversion formula leading to an
explicit formula for $u$.
Eq.\,(\ref{Eq:I_23}) simplifies considerably, if we set
$\Omega_{\hbox{\scriptsize r}}\equiv 0$, i.\,e.\ neglect radiation.
We then have to solve the integral
\begin{equation}
\label{Eq:I_23_Or_0}
\left. I_{-\frac 23}(u) \right|_{\Omega_{\hbox{\scriptsize r}}\equiv 0} \; = \;
\int_0^u \frac{1}
{\sqrt{\Omega_\phi y^2 + \Omega_{\hbox{\scriptsize c}} y +
\Omega_{\hbox{\scriptsize m}}}} \; dy
\hspace{10pt} ,
\end{equation}
which is identical to the integral $I_{-\frac 13}(u)$,
if we make the replacements
$\tilde \Omega_\phi = \Omega_\phi+\Omega_{\hbox{\scriptsize c}} \to
\Omega_\phi$,
$\Omega_{\hbox{\scriptsize m}} \to \Omega_{\hbox{\scriptsize c}}$ and
$\Omega_{\hbox{\scriptsize r}} \to \Omega_{\hbox{\scriptsize m}}$.
We can thus immediately obtain the corresponding scale factor by
making the above replacements in formula (\ref{Eq:A_13}) and then obtain
\begin{equation}
\label{Eq:scale_factor_eta_check}
\frac{A(\phi)}{a_0} \; = \;
\frac 12 \, \frac{\Omega_{\hbox{\scriptsize c}}}{\Omega_\phi} \,
\left( \check \eta \sinh(B\phi) + \cosh(B\phi) - 1 \right)
\end{equation}
with $\check \eta := 2 \frac{\sqrt{\Omega_\phi\Omega_{\hbox{\scriptsize m}}}}
{\Omega_{\hbox{\scriptsize c}}}$ and
$B = \frac{2\sqrt{2\pi}}{m_P}$.
Inserting (\ref{Eq:scale_factor_eta_check}) into our general formula
(\ref{Eq:potential_A}), gives then the
{\it exact quintessence potential} for a two-component model consisting
of matter and quintessence  with $w_\phi=-\frac 23$
\begin{equation}
\label{Eq:V_23_matter_phi}
V(\phi) \; = \; \frac{\check V_0}
{\check \eta \sinh(B\phi) + \cosh(B\phi) - 1}
\end{equation}
with $\check V_0 := \frac 53 \frac{\Omega_\phi^2}
{\Omega_{\hbox{\scriptsize c}}} \varepsilon_{\hbox{\scriptsize crit}}$.
Explicitly, one obtains
\begin{equation}
\label{Eq:V_23_matter_phi_limit}
V(\phi) \; = \;
\left\{ \begin{matrix} 
\frac{\check V_0}{\check \eta} \, \, \frac 1{B\phi}
\hbox{ for } \phi \to 0 \\
\frac{2\check V_0}{1+\check \eta} \, e^{-B\phi}
\hbox{ for } \phi \to \infty
\end{matrix} \right.
\hspace{10pt} ,
\end{equation}
which is in accordance with the general behavior given by
Eqs.(\ref{Eq:Pot_limit_0}), (\ref{Eq:V_limit_infty}) and (\ref{Def:mu}).

Since radiation has been neglected in the derivation of the
potential (\ref{Eq:V_23_matter_phi}), it appears at first sight
that (\ref{Eq:V_23_matter_phi}) can be applied only in the
matter-dominated epoch and later,
i.\,e.\ for $\eta > \eta_{\hbox{\scriptsize m}}$ respectively
$z < z_{\hbox{\scriptsize m}}$ (see Eq.\,(\ref{Eq:z_m})).
We can consider, however, a model in which the potential
(\ref{Eq:V_23_matter_phi}) is assumed to govern the time-evolution
of the quintessence field for all times, irrespective of its derivation,
and solve Eqs.\,(\ref{Eq:eos_phi})$-$(\ref{Eq:continuity_equation})
in a full three-component background model consisting of radiation, matter
and quintessence.
Radiation is then included correctly by means of the scale factor $a(\eta)$,
and thus the new model is consistent for all times.
It is clear, however, that the equation of state $w_\phi(\eta)$
can no more be constant for all $\eta \geq 0$.
Nevertheless, it is obvious that for the potential (\ref{Eq:V_23_matter_phi})
asymptotically holds
\begin{equation}
\label{Eq:w_23_matter_phi_limit}
w_\phi(\eta) \; \to \; -\frac 23
\hspace{10pt} \hbox{ for } \hspace{10pt}
\eta \; \to \; \eta_\infty
\hspace{10pt} ,
\end{equation}
where we expect $w_\phi(\eta)$ to be {\it almost} constant,
$w_\phi(\eta) \simeq -\frac 23$, already for
$\eta \gtrsim \eta_{\hbox{\scriptsize m}}$.
But in the radiation-dominated epoch
$0\leq \eta \leq \eta_{\hbox{\scriptsize m}}$,
the equation of state will deviate from its asymptotic value $-\frac 23$.
It is then interesting to study the behavior of $w_\phi(\eta)$ in the
limit $\eta\to 0$.

For this purpose, let us consider the more general situation of
a quintessence potential $V(\phi)$ for a two-component model
consisting of matter and quintessence with a constant equation of state
$w_\infty$, $-1 < w_\infty\leq -\frac 13$.
It follows from Eqs.\,(\ref{Eq:phi_sing}) and (\ref{Eq:Pot_limit_0})
that $V(\phi)$ diverges in the radiation-dominated epoch like
\begin{equation}
\label{Eq:V_divergence_w_infty}
V(\phi) \; = \; \frac{\overline{V}_0}{\phi^{\overline{\gamma}}} \, + \, \dots
\hspace{10pt} , \hspace{10pt}
\overline{\gamma} \; = \; - \, \frac 2{w_\infty}(1+w_\infty)
\hspace{10pt} .
\end{equation}
We then assume, as in the discussion above, that this potential is taken
as a model for quintessence using, however,
in Eqs.\,(\ref{Eq:eos_phi})$-$(\ref{Eq:continuity_equation})
the scale factor $a(\eta)$ for a three-component model
which takes also radiation into account.
One then obtains a time dependent equation of state, $w_\phi(\eta)$,
which asymptotically obeys
\begin{equation}
\label{Eq:w_asymptotic}
w_\phi(\eta) \; \to \; w_\infty
\hspace{10pt} , \hspace{10pt}
\eta \; \to \;\eta_\infty
\hspace{10pt} .
\end{equation}
(We expect $w_\phi(\eta) \simeq w_\infty$ to hold for
$\eta \gtrsim \eta_{\hbox{\scriptsize m}}$.)
To derive the behavior of $w_\phi(\eta)$ in the opposite limit,
$\eta \to 0$, we make the ansatz $\phi(\eta) = b \eta^\sigma + \dots$
for $\eta \to 0$ with $b>0$, $\sigma>0$.
Since the background model includes radiation, we have
$a(\eta) = a'(0) \eta + \dots$ in this limit.
With (\ref{Eq:V_divergence_w_infty}), it is not difficult to see
that the equation of motion (\ref{Eq:equation_of_motion_phi})
is satisfied iff $\sigma = 4/(\overline{\gamma}+2)$ and
$b^{\overline{\gamma}+2} = \overline{\gamma} {a'}^2(0)\overline{V}_0/
(\sigma(\sigma+1))$.
It then follows for the ratio of the kinetic to the potential energy
(see Eq.(\ref{Eq:w_r}))
$r_\phi(\eta) \to r_\phi(0)$, $\eta \to 0$, with
$$
r_\phi(0) \; = \;
\frac{b^{\overline{\gamma}+2} \sigma^2}{2{a'}^2(0)\overline{V}_0}
\; = \;
\frac{2\overline{\gamma}}{\overline{\gamma}+6}
$$
which yields
\begin{equation}
\label{Eq:w_initial}
w_\phi(\eta) \; \to \; w_\phi(0) \, = \, \frac 13 (1+4w_\infty)
\hspace{10pt} , \hspace{10pt}
\eta \to 0
\hspace{10pt} .
\end{equation}
(Notice that $-1 < w_\phi(0) \leq - \frac 19$ and $w_\phi(0)>w_\infty$
for $-1 < w_\infty \leq - \frac 13$.)
We thus conclude that the equation of state for the above models starts with
the initial value (\ref{Eq:w_initial}) at $\eta=0$ and then decreases to its
asymptotic value $w_\infty$.
In the special case $w_\infty=-\frac 23$, discussed at the beginning of
this Section, we have $w_\phi(0)=-\frac 59$, and for the potential
(\ref{Eq:V_13_without_radiation}) with $w_\infty=-\frac 13$
one gets $w_\phi(0)=-\frac 19$.

\subsection{Potentials with a Positive Cosmological Constant}
\label{Potentials_with_Cosmological_Constant}

\subsubsection{General Properties}

In the foregoing discussion, we have considered a three-component
model consisting of radiation, matter and quintessence.
We will now study a four-component model, which takes into account also
a non-vanishing (positive) cosmological constant $\Lambda$
corresponding to an energy density (vacuum energy density)
\begin{equation}
\label{Def:epsilon_Lambda}
\varepsilon_\Lambda \; := \; 
\frac{\Lambda}{8\pi G} \; = \; \hbox{const.}
\end{equation}
with density parameter $\Omega_\Lambda = \frac{\Lambda}{3H_0^2}>0$.
Assuming $-1 < w_\phi \leq -\frac 13$, we have now to compute instead
of (\ref{Def:I_w}) the more general integral
\begin{equation}
\label{Def:I_w_Lambda}
\tilde I_w(u) \: := \:
\int_0^u \frac{y^{-\frac 12 - \frac 32 w} \; dy}
{\sqrt{\Omega_\Lambda y^4 + \Omega_\phi y^{1-3w} +
\Omega_{\hbox{\scriptsize c}} y^2 + \Omega_{\hbox{\scriptsize m}} y +
\Omega_{\hbox{\scriptsize r}}}}
\hspace{5pt} .
\end{equation}
Let us begin with some general remarks.
Obviously, the leading asymptotic behavior of $\tilde I_w(u)$ in the
limit $u\to 0$ is the same as for $I_w(u)$, Eq.\,(\ref{Def:I_w})
(see also (\ref{Eq:phi_sing})), and thus we obtain in the radiation-dominated
epoch again the power-law divergence (\ref{Eq:Pot_limit_0}) with the
same power $\gamma$ as in the case of the three-component models.
In the limit $u\to \infty$, however, we get a completely different behavior,
which follows from the fact that $\tilde I_w(u)$, Eq.\,(\ref{Def:I_w_Lambda}),
stays finite in this limit.
(Indeed, the integrand of $\tilde I_w(u)$ behaves for $y\to \infty$ like
$y^{-\frac 52 - \frac 32 w}$, and thus the integral (\ref{Def:I_w_Lambda})
converges at the upper limit since $\frac 52 + \frac 32 w>1$ for $w>-1$.)
Similarly, it follows that the corresponding generalization of the integral
(\ref{Eq:eta_a}) stays finite in the limit $a(\eta)\to\infty$,
which implies that the conformal time $\eta$ approaches at late times
the finite value $\eta_\infty$ given by
\begin{equation}
\label{Eq:eta_infty_Lambda}
\eta_\infty =
\sqrt{\Omega_{\hbox{\scriptsize c}}} \int_0^\infty
\hspace*{-8pt}\frac{dy}{\sqrt{\Omega_\Lambda y^4 + \Omega_\phi y^{1-3w_\phi} +
\Omega_{\hbox{\scriptsize c}} y^2 + \Omega_{\hbox{\scriptsize m}} y +
\Omega_{\hbox{\scriptsize r}} }}
\hspace{2pt} .
\end{equation}
We thus conclude that the quintessence field $\phi(\eta)$ approaches
in the limit $\eta\to \eta_\infty$ the finite value
$\phi_\infty := \phi(\eta_\infty)$ given by
(see Eqs.\,(\ref{Eq:phi_final}) and (\ref{Def:B_0}))
\begin{equation}
\label{Eq:phi_final_Lambda}
\phi_\infty = \frac 1{B_0} \int_0^\infty
\hspace*{-8pt}\frac{y^{-\frac 12 - \frac 32 w_\phi} \;dy}
{\sqrt{\Omega_\Lambda y^4 + \Omega_\phi y^{1-3w_\phi} +
\Omega_{\hbox{\scriptsize c}} y^2 + \Omega_{\hbox{\scriptsize m}} y +
\Omega_{\hbox{\scriptsize r}} }}
\hspace{2pt} .
\end{equation}
With (\ref{Eq:phi_final_Lambda}), the generalization of
Eq.\,(\ref{Eq:phi_final})
can be rewritten
\begin{equation}
\label{Eq:B_0_Lambda}
B_0 \, \phi \; = \;
B_0 \, \phi_\infty - \hat I_{w_\phi}\left(\frac{A(\phi)}{a_0}\right)
\hspace{10pt} ,
\end{equation}
where we have defined the function
\begin{equation}
\label{Def:I_hat}
\hat I_w(u) := \int_u^\infty
\frac{y^{-\frac 12 - \frac 32 w} \;dy}
{\sqrt{\Omega_\Lambda y^4 + \Omega_\phi y^{1-3w} +
\Omega_{\hbox{\scriptsize c}} y^2 + \Omega_{\hbox{\scriptsize m}} y +
\Omega_{\hbox{\scriptsize r}} }}
\hspace{5pt} .
\end{equation}
In the limit $u\to\infty$, one obtains for $w>-1$
\begin{equation}
\label{Eq:I_hat_limit}
\hat I_w(u) = \frac 23 \, \frac 1{(1+w) \sqrt{\Omega_\Lambda}}
\, u^{-\frac 32(1+w)} \, + \, \dots
\hspace{10pt} ,
\end{equation}
which yields together with (\ref{Eq:B_0_Lambda}) the following asymptotic
behavior of the scale factor $A(\phi)$
\begin{equation}
\label{Eq:A_limit}
\frac{A(\phi)}{a_0} \; = \;
O\left( (\phi_\infty-\phi)^{-\frac 2{3(1+w_\phi)}}\right)
\hspace{10pt} , \hspace{10pt}
\phi \to \phi_\infty
\hspace{10pt} .
\end{equation}
Inserting the last result in our general formula (\ref{Eq:potential_A}),
gives for $\Lambda>0$, $w_\phi>-1$
\begin{equation}
\label{Eq:V_limit_Lambda}
V(\phi) \; = \;
3 \pi (1-w_\phi^2) \Omega_\Lambda \varepsilon_{\hbox{\scriptsize crit}}
\left( \frac{\phi_\infty-\phi}{m_P}\right)^2  \, + \, \dots
\hspace{2pt} , \hspace{2pt}
\phi \to \phi_\infty
\hspace{1pt} .
\end{equation}
Thus, the quintessence potential shows at late times $(\eta\to\eta_\infty)$
in a universe with a positive cosmological constant a completely
different behavior from that exhibited by Eq.\,(\ref{Eq:V_limit_infty}).
While $V(\phi)$ vanishes in the case $\Lambda\equiv 0$ exponentially
for $\phi\to\infty$, it vanishes for $\Lambda>0$ only like a power,
Eq.\,(\ref{Eq:V_limit_Lambda}), at a finite value $\phi_\infty$
for $\phi\to\phi_\infty$.

In the case $\Lambda >0$, we have to replace Eq.\,(\ref{Eq:Omega_redshift_g})
for the time-evolution of the density parameters by
\begin{equation}
\label{Eq:Omega_redshift_g_Lambda}
\Omega_{\hbox{\scriptsize x}}(\eta) \; = \;
\frac{y^{1-3w_{\hbox{\scriptsize x}}}}{g_\Lambda(y)} \,
\Omega_{\hbox{\scriptsize x}}
\hspace{10pt} , \hspace{10pt}
\hbox{x} = \{\hbox{r}, \hbox{m},\phi,\Lambda\}
\end{equation}
and Eq.\,(\ref{Eq:Omega_tot_limit_0}) for
$\Omega_{\hbox{\scriptsize tot}}(\eta)$ by
\begin{equation}
\label{Eq:Omega_tot_limit_0_Lambda}
\Omega_{\hbox{\scriptsize tot}}(\eta) \; = \;
1 \, - \, \frac{y^2}{g_\Lambda(y)} \, (1-\Omega_{\hbox{\scriptsize tot}})
\end{equation}
with (see Eq.(\ref{Def:g}))
\begin{equation}
\label{Def:g_Lambda}
g_\Lambda(y) \; := \; \Omega_\Lambda y^4 \, + \, g(y) \; = \;
\sum_{\hbox{\scriptsize x}=\hbox{\scriptsize r},
\hbox{\scriptsize m},\phi,\Lambda,\hbox{\scriptsize c}}
\Omega_{\hbox{\scriptsize x}} \, y^{1-3w_{\hbox{\scriptsize x}}}
\hspace{10pt} .
\end{equation}
Obviously, the behavior (\ref{Eq:Omega_eta_small}) of the density
parameters in the limit $\eta\to 0$ is not changed, and we only have to add
$\Omega_\Lambda(\eta) = O(\eta^4)$ for $\eta\to 0$.
At late times, $z\to-1$, however, Eqs.\,(\ref{Eq:Omega_eta_large_13})
and (\ref{Eq:Omega_eta_large}) have to be replaced by
$(w_\phi>-1)$
\begin{equation}
\label{Eq:Omega_eta_large_Lambda}
\begin{matrix}
\Omega_{\hbox{\scriptsize r}}(\eta) & = &
\frac{\Omega_{\hbox{\scriptsize r}}}{\Omega_\Lambda} \, (1+z)^4
\, + \, \dots \; \to \; 0  \\
\Omega_{\hbox{\scriptsize m}}(\eta) & = & 
\frac{\Omega_{\hbox{\scriptsize m}}}{\Omega_\Lambda} \, (1+z)^3
\, + \, \dots \; \to \; 0 \\
\Omega_\phi(\eta) & = &
\frac{\Omega_\phi}{\Omega_\Lambda} \, (1+z)^{3+3w_\phi}
\, + \, \dots \; \to \; 0 \\
\Omega_\Lambda(\eta) & = &
1 - \frac{\Omega_\phi}{\Omega_\Lambda} \, (1+z)^{3+3w_\phi}
+ \dots \to 1 \\
\Omega_{\hbox{\scriptsize tot}}(\eta) & = &
1 - \frac{\Omega_{\hbox{\scriptsize c}}}{\Omega_\Lambda} \, (1+z)^2
+ \dots \to 1
\end{matrix}  \hspace{10pt} .
\end{equation}
We see that in the case $\Lambda>0$ also the quintessence component
vanishes asymptotically at late times, in contrast to the earlier
behavior (\ref{Eq:Omega_eta_large}), since now the vacuum contribution
$\Omega_\Lambda(\eta)$ dominates and approaches one in this limit.
Furthermore, we observe that for $\Lambda>0$ the universe becomes
asymptotically flat.
Quintessence-vacuum energy equality holds at $z=z_{\phi\Lambda}$ with
\begin{equation}
\label{Eq:z_phi_Lambda}
z_{\phi\Lambda} \; = \; \left(\frac{\Omega_\Lambda}
{\Omega_\phi}\right)^{\frac 1{3+3w_\phi}} \, - \, 1
\hspace{10pt} .
\end{equation}

\subsubsection{The Potential for $w_\phi=-\frac 13$ and $\Lambda>0$}

For $w_\phi=-\frac 13$ and $\Lambda>0$, we have
(see Eqs.\,(\ref{Eq:phi_final}) and (\ref{Def:I_w_Lambda}))
\begin{equation}
\label{Eq:B_0_Lambda_int}
B_0 \, \phi \; = \;
\tilde I_{-\frac 13}\left(\frac{A(\phi)}{a_0}\right)
\hspace{10pt} ,
\end{equation}
and thus we have to ``invert'' the integral
\begin{equation}
\label{Eq:integ_invert}
\zeta \; := \; \tilde I_{-\frac 13}(u) \; = \;
\int_0^u \frac{dy}
{\sqrt{\Omega_\Lambda y^4 + \tilde\Omega_\phi y^2 +
\Omega_{\hbox{\scriptsize m}} y + \Omega_{\hbox{\scriptsize r}} }}
\hspace{10pt} .
\end{equation}
The integral (\ref{Eq:integ_invert}) is an elliptic integral
of the first kind, which is insoluble by means of elementary functions.
However, it is possible to invert it and to express $u$ as a rational
function of ${\cal P}(\zeta)$, where ${\cal P}(\zeta)$ is the Weierstrass
${\cal P}$-function.
It then follows from (\ref{Eq:B_0_Lambda_int}) that $A(\phi)/a_0$
can be expressed as a rational function of ${\cal P}(B_0\phi)$.
(This method has already been used in our earlier paper
\cite{Aurich_Steiner_2000} to express $a(\eta)$ in terms of ${\cal P}(\eta)$).

Let $F(y) := a_0y^4 + 4a_1 y^3 + 6 a_2 y^2 + 4 a_3 y + a_4$ be any
quartic polynomial which has no repeated factors;
and let its {\it invariants} be
\cite{Whittaker_Watson_1973}
\begin{eqnarray} \nonumber
g_2 & := & a_0 a_4 - 4 a_1 a_3 + 3 a_2^2 
\\ g_3 & := & \label{Eq:Weierstrass_invariants}
a_0 a_2 a_4 + 2 a_1 a_2 a_3 - a_2^3 -a_0 a_3^2 - a_1^2 a_4
\hspace{10pt} .
\end{eqnarray}
Furthermore, let
\begin{equation}
\label{Eq:zeta_integral}
\zeta \; = \; \int_\alpha^u \frac{dy}{\sqrt{F(y)}}
\hspace{10pt} ,
\end{equation}
where $\alpha$ is any constant.
(In the applications, which we have in mind, $\alpha$ is real and non-negative
and $F(y)>0$ for $y\geq \alpha$.)
Then \cite{Whittaker_Watson_1973}
\begin{eqnarray}\nonumber
u & = & u(\zeta) = \alpha +
\frac{\frac 12 F'(\alpha)
\left({\cal P}(\zeta) - \frac 1{24}F''(\alpha)\right)}
{2\left({\cal P}(\zeta) - \frac 1{24}F''(\alpha)\right)^2 -
\frac 1{48}F(\alpha)F^{(4)}(\alpha)}
\\ & & \label{Eq:u_alpha} \hspace*{10pt} + \;
\frac{\frac 1{24}F(\alpha) F'''(\alpha)-\sqrt{F(\alpha)} {\cal P}'(\zeta)}
{2\left({\cal P}(\zeta) - \frac 1{24}F''(\alpha)\right)^2 -
\frac 1{48}F(\alpha)F^{(4)}(\alpha)}
\hspace{5pt} ,
\end{eqnarray}
the Weierstrass function ${\cal P}(\zeta) = {\cal P}(\zeta;g_2,g_3)$
being formed with the invariants (\ref{Eq:Weierstrass_invariants})
of the quartic $F(y)$.
(Notice that we have corrected a sign on p.\,454 in
\cite{Whittaker_Watson_1973} by writing $-\sqrt{F(\alpha)} {\cal P}'(\zeta)$
in Eq.\,(\ref{Eq:u_alpha}).)
${\cal P}(\zeta)$ can numerically be evaluated very efficiently by
its Laurent expansion
\begin{equation}
\label{Eq:Weierstrass}
{\cal P}(\zeta) \; = \;  \frac 1{\zeta^2} \, + \,
\sum_{k=2}^\infty c_k \zeta^{2k-2}
\end{equation}
with
$$
c_2 := \frac{g_2}{20}
\hspace{10pt} , \hspace{10pt} c_3 := \frac{g_3}{28}
$$
and the recursion relation \cite{Abramowitz_Stegun_1965}
$$
c_k = \frac{3}{(2k+1)(k-3)} \sum_{m=2}^{k-2} c_m c_{k-m} \; \hbox{ for
} \; k\geq 4
\hspace{10pt} .
$$
With $a_0=\Omega_\Lambda$, $a_1=0$,
$a_2=(\Omega_\phi+\Omega_{\hbox{\scriptsize c}})/6$,
$a_3 = \Omega_{\hbox{\scriptsize m}}/4$,
$a_4 = \Omega_{\hbox{\scriptsize r}}$ and $\alpha=0$,
we obtain from Eqs.\,(\ref{Eq:B_0_Lambda_int}), (\ref{Eq:integ_invert}),
(\ref{Eq:zeta_integral}) and (\ref{Eq:u_alpha})
\begin{equation}
\label{Eq:A_Weierstrass}
\frac{A(\phi)}{a_0} \; = \; \frac{\Omega_{\hbox{\scriptsize m}}}4
\frac{{\cal P}(B_0\phi) - \frac{\Omega_\phi+\Omega_{\hbox{\scriptsize c}}}{12}
- 2 \frac{\sqrt{\Omega_{\hbox{\scriptsize r}}}}{\Omega_{\hbox{\scriptsize m}}}
{\cal P}'(B_0\phi)}
{\left({\cal P}(B_0\phi) - \frac{\Omega_\phi+\Omega_{\hbox{\scriptsize c}}}{12}
\right)^2 - \frac{\Omega_{\hbox{\scriptsize r}}\Omega_\Lambda}4}
\end{equation}
with $B_0 = \frac{2\sqrt \pi}{m_P} \Omega_\phi^{-1/2}$,
where ${\cal P}(B_0\phi;g_2,g_3)$ has to be formed with the {\it invariants}
\begin{eqnarray}
\label{Eq:Weierstrass_invariants_Omega}
g_2 & := & \frac{\left(\Omega_\phi+\Omega_{\hbox{\scriptsize c}}\right)^2}{12}
+ \Omega_{\hbox{\scriptsize r}}\Omega_\Lambda
\\ g_3 & := &  \nonumber
- \frac{\left(\Omega_\phi+\Omega_{\hbox{\scriptsize c}}\right)^3}{216}
+ \left(\frac{\Omega_{\hbox{\scriptsize r}}
\left(\Omega_\phi+\Omega_{\hbox{\scriptsize c}}\right)}6 -
\frac{\Omega_{\hbox{\scriptsize m}}^2}{16} \right) \Omega_\Lambda
\hspace{10pt} .
\end{eqnarray}
Inserting (\ref{Eq:A_Weierstrass}) into our general formula
(\ref{Eq:potential_A}), yields the {\it exact quintessence potential}
for a four-component model with $\Lambda>0$ and $w_\phi=-\frac 13$
\begin{eqnarray}
\label{Eq:V_13_Lambda}
V(\phi) & = &
\frac{4V_0}{\left(\Omega_\phi+\Omega_{\hbox{\scriptsize c}}\right)^2} \,
\\ & & \nonumber \times \;
\left[\frac{\left({\cal P}(B_0\phi) -
\frac{\Omega_\phi+\Omega_{\hbox{\scriptsize c}}}{12}\right)^2 -
\frac{\Omega_{\hbox{\scriptsize r}}\Omega_\Lambda}4}
{{\cal P}(B_0\phi) -
\frac{\Omega_\phi+\Omega_{\hbox{\scriptsize c}}}{12} -
2\frac{\sqrt{\Omega_{\hbox{\scriptsize r}}}}{\Omega_{\hbox{\scriptsize m}}}
{\cal P}'(B_0\phi)}
\right]^2
\hspace{10pt} ,
\end{eqnarray}
where the potential strength $V_0$ is found to be the same as in
Eq.(\ref{Eq:V_13}).
Using the leading behavior
${\cal P}(B_0\phi) = (B_0\phi)^{-2} + \dots$ and
${\cal P}'(B_0\phi) = -2(B_0\phi)^{-3} + \dots$ for $\phi\to 0$,
it follows that (\ref{Eq:V_13_Lambda}) obeys the correct
power-law behavior (\ref{Eq:Pot_limit_0})
(see also Eq.\,(\ref{Eq:V_13_limit})) in the radiation-dominated epoch
$\phi\to 0$.
To derive the asymptotic behavior of $V(\phi)$ in the opposite limit at
late times, $\eta\to\eta_\infty$ or $\phi\to\phi_\infty$,
(see Eqs.\,(\ref{Eq:eta_infty_Lambda}) and (\ref{Eq:phi_final_Lambda})),
we use \cite{Whittaker_Watson_1973}
$$
{\cal P}(\zeta) \; = \;
\frac{\sqrt{F(u)F(\alpha)}+F(\alpha)}{2(u-\alpha)^2} +
\frac{F'(\alpha)}{4(u-\alpha)} + \frac{F''(\alpha)}{24}
$$
\begin{eqnarray}
\label{Eq:Weierstrass_F}
{\cal P}'(\zeta) & = &
- \left[\frac{F(u)}{(u-\alpha)^3} - \frac{F'(u)}{4(u-\alpha)^2}
\right] \sqrt{F(\alpha)}
\\ & & \nonumber
- \left[\frac{F(\alpha)}{(u-\alpha)^3} + \frac{F'(\alpha)}{4(u-\alpha)^2}
\right] \sqrt{F(u)}
\end{eqnarray}
with the same notation as in Eq.\,(\ref{Eq:zeta_integral}).
From Eqs.\,(\ref{Eq:B_0_Lambda_int}) and (\ref{Eq:integ_invert}) we infer
that we have to choose $\alpha=0$ in Eq.\,(\ref{Eq:Weierstrass_F})
and then consider the limit $u=A(\phi)/a_0\to \infty$, which yields
\begin{eqnarray} \nonumber
{\cal P}(B_0\phi_\infty) & = &
\frac{\sqrt{F(0)}}2 \, \lim_{u\to\infty} \, \frac{\sqrt{F(u)}}{u^2} +
\frac{F''(0)}{24}
\\ & = &
\label{Eq:Weierstrass_Omega}
\frac{\sqrt{\Omega_{\hbox{\scriptsize r}}\Omega_\Lambda}}2 +
\frac{\Omega_\phi+\Omega_{\hbox{\scriptsize c}}}{12}
\end{eqnarray}
and
\begin{eqnarray} \nonumber
{\cal P}'(B_0\phi_\infty) & = &
- \sqrt{F(0)} \, \lim_{u\to\infty} \, \left[
\frac{F(u)}{u^3} - \frac{F'(u)}{4u^2} \right]
\\ & & \nonumber
- \frac{F'(0)}4 \, \lim_{u\to\infty} \, \frac{\sqrt{F(u)}}{u^2}
\\ & = &
\label{Eq:Weierstrass_Omega_prime}
- \, \frac{\Omega_{\hbox{\scriptsize m}}}4 \sqrt{\Omega_\Lambda}
\hspace{10pt} .
\end{eqnarray}
The implicit relation (\ref{Eq:Weierstrass_Omega}) turns out to be
very convenient to calculate $\phi_\infty$ numerically
(instead of computing the integral (\ref{Eq:phi_final_Lambda})).
A Taylor expansion of ${\cal P}(B_0\phi)$ and
${\cal P}'(B_0\phi)$ at $\phi=\phi_\infty$ gives then for the
potential (\ref{Eq:V_13_Lambda}) the desired result in the
limit $\phi\to\phi_\infty$.
Explicitly, one derives
\begin{equation}
\label{Eq:V_13_Lambda_limit}
V(\phi) \; = \;
\left\{ \begin{matrix} 
\frac{V_0\Omega_{\hbox{\scriptsize m}}^2}{16\Omega_{\hbox{\scriptsize r}}}
\, \frac 1{(B_0\phi)^2}
\hbox{ for } \phi \to 0 \\
\frac 23 \varepsilon_{\hbox{\scriptsize crit}} \Omega_\phi \Omega_\Lambda
\left( B_0(\phi_\infty-\phi) \right)^2
\hbox{ for } \phi \to \phi_\infty
\end{matrix} \right.
\end{equation}
in full agreement with the general behavior (\ref{Eq:Pot_limit_0}) and
(\ref{Eq:V_limit_Lambda}).

It is worthwhile to check whether the potential (\ref{Eq:V_13_Lambda})
goes in the limit of a vanishing cosmological constant,
$\Omega_\Lambda\to 0$, over to our previous result (\ref{Eq:V_13}).
From (\ref{Eq:Weierstrass_invariants_Omega}), we obtain for the
invariants in this limit
$g_2 \to (\Omega_\phi+\Omega_{\hbox{\scriptsize c}})^2/12$ and
$g_3 \to -(\Omega_\phi+\Omega_{\hbox{\scriptsize c}})^3/216$.
Using the homogeneity property of the function ${\cal P}(\zeta)$
\cite{Whittaker_Watson_1973},
${\cal P}(\zeta;g_2,g_3) =
\lambda^2{\cal P}(\lambda\zeta;\frac{g_2}{\lambda^4},\frac{g_3}{\lambda^6})$,
we obtain with $\lambda=\sqrt{\Omega_\phi+\Omega_{\hbox{\scriptsize c}}}$
\begin{equation}
\label{Eq:Weierstrass_vanishing_Lambda}
{\cal P}(B_0\phi;g_2,g_3) \; \to_{(\Omega_\Lambda\to 0)} \;
(\Omega_\phi+\Omega_{\hbox{\scriptsize c}})\,
{\cal P}(B\phi;\frac 1{12}, -\frac 1{216})
\hspace{5pt} ,
\end{equation}
where we have used the definition
$B := B_0\sqrt{\Omega_\phi+\Omega_{\hbox{\scriptsize c}}}$,
see Eq.\,(\ref{Def:B}).
Since the last ${\cal P}$-function can be expressed in terms of an
elementary function \cite{Abramowitz_Stegun_1965},
\begin{equation}
\label{Eq:Weierstrass_elementary_function}
{\cal P}\left(\zeta;\frac 1{12}, -\frac 1{216}\right) \; = \;
\frac 1{12} \, + \, \frac 1{4\sinh^2\frac{\zeta}2}
\hspace{10pt} ,
\end{equation}
one obtains, indeed, that the potential (\ref{Eq:V_13_Lambda})
goes over to the potential (\ref{Eq:V_13})
if the cosmological constant approaches zero.

\subsubsection{The Potential for $w_\phi=-\frac 23$, $\Lambda>0$
and $\Omega_{\hbox{\scriptsize r}}\equiv 0$}

For $w_\phi=-\frac 23$, we have to solve
(see Eqs.\,(\ref{Eq:B_0_Lambda}) and (\ref{Def:I_hat}))
\begin{equation}
\label{Eq:B_0_23}
B_0(\phi_\infty -\phi) \; = \;
\hat I_{-\frac 23}\left(\frac{A(\phi)}{a_0}\right)
\end{equation}
with
\begin{equation}
\label{Def:I_hat_23}
\hat I_{-\frac 23}(u) = \int_u^\infty
\frac{\sqrt{y} \;dy}
{\sqrt{\Omega_\Lambda y^4 + \Omega_\phi y^3 +
\Omega_{\hbox{\scriptsize c}} y^2 + \Omega_{\hbox{\scriptsize m}} y +
\Omega_{\hbox{\scriptsize r}} }}
\hspace{10pt} .
\end{equation}
The last integral simplifies considerably, if we set $\Omega_r\equiv 0$,
i.\,e.\ neglect radiation.
Introducing the new variable of integration $t$,
$y := \frac 4{\Omega_\Lambda} \left( t- \frac{\Omega_\phi}{12} \right)$,
the integrand is transformed into Weierstrass normal form
\begin{equation}
\label{Eq:B_normal_23}
B_0(\phi_\infty-\phi) \; = \;
\int_T^\infty \frac{dt}{\sqrt{4t^3-g_2 t -g_3}}
\end{equation}
with {\it invariants}
\begin{eqnarray}
\label{Eq:Weierstrass_invariants_Omega_23}
g_2 & := & \frac{\Omega_\phi^2}{12} -
\frac{\Omega_{\hbox{\scriptsize c}}\Omega_\Lambda} 4
\\ g_3 & := &  \nonumber
- \frac{\Omega_\phi^3}{216} + \frac 1{48} \left(
\Omega_{\hbox{\scriptsize c}}\Omega_\phi -
3 \Omega_{\hbox{\scriptsize m}}\Omega_\Lambda\right) \Omega_\Lambda
\end{eqnarray}
and $T := \frac{\Omega_\Lambda}4 \left(\frac{A(\phi)}{a_0}\right) +
\frac{\Omega_\phi}{12}$.
In the form (\ref{Eq:B_normal_23}), the integral can be solved
\cite{Whittaker_Watson_1973} in terms of the Weierstrass
${\cal P}$-function, $T = {\cal P}(B_0(\phi_\infty-\phi))$,
and one immediately obtains for the scale factor
\begin{equation}
\label{Eq:scale_factor_23}
\frac{A(\phi)}{a_0} \; = \;
\frac 4{\Omega_\Lambda} \, {\cal P}(B_0(\phi_\infty-\phi)) \, - \,
\frac 13 \, \frac{\Omega_\phi}{\Omega_\Lambda}
\hspace{10pt} ,
\end{equation}
where the function ${\cal P}$ has to be formed with the invariants
(\ref{Eq:Weierstrass_invariants_Omega_23}).
Inserting (\ref{Eq:scale_factor_23}) into our general formula 
(\ref{Eq:potential_A}), yields the {\it exact quintessence potential}
for a three-component model consisting of matter, quintessence and
a cosmological constant
with $w_\phi=-\frac 23$, $\Omega_\Lambda>0$ and
$\Omega_{\hbox{\scriptsize r}}\equiv 0$
\begin{equation}
\label{Eq:V_23_Lambda}
V(\phi) \; = \;
\frac{\hat V_0}{{\cal P}(B_0(\phi_\infty-\phi)) - \frac{\Omega_\phi}{12}}
\end{equation}
with $\hat V_0 = \frac 5{24} \varepsilon_{\hbox{\scriptsize crit}}
\Omega_\phi \Omega_\Lambda$.
The potential (\ref{Eq:V_23_Lambda}) depends on the field $\phi_\infty$,
which is defined by
\begin{eqnarray} \nonumber
\phi_\infty & = & \label{Eq:phi_infty_23}
\frac 1{B_0} \int_0^\infty
\frac{dy}
{\sqrt{\Omega_\Lambda y^3 + \Omega_\phi y^2 +
\Omega_{\hbox{\scriptsize c}} y + \Omega_{\hbox{\scriptsize m}}}}
\\ & = & 
\frac 1{B_0} \int_{\Omega_\phi/12}^\infty \frac{dt}{\sqrt{4t^3-g_2 t -g_3}}
\end{eqnarray}
and can be computed from
\begin{equation}
\label{Eq:P_B_phi_infty}
{\cal P}(B_0\phi_\infty) \; = \; \frac{\Omega_\phi}{12}
\hspace{10pt} .
\end{equation}
Expanding ${\cal P}(B_0(\phi_\infty-\phi))$ into a Taylor series at
$\phi=0$ and using (\ref{Eq:P_B_phi_infty}) and
\begin{equation}
{\cal P}'(B_0\phi_\infty) \; = \;
- \, \frac{\sqrt{\Omega_{\hbox{\scriptsize m}}}}4 \, \Omega_\Lambda
\hspace{10pt} ,
\end{equation}
one derives the asymptotic behavior of the potential (\ref{Eq:V_23_Lambda})
for $\phi\to 0$.
The behavior of $V(\phi)$ in the limit $\phi\to\phi_\infty$ follows
immediately form (\ref{Eq:Weierstrass}).
Explicitly, one obtains
\begin{equation}
\label{Eq:V_23_Lambda_limit}
V(\phi) \; = \;
\left\{ \begin{matrix} 
\frac{4 \hat V_0}{\sqrt{\Omega_{\hbox{\scriptsize m}}}\Omega_\Lambda}
\, \frac 1{B_0\phi}
\hbox{ for } \phi \to 0 \\
\hat V_0 \, \left( B_0(\phi_\infty-\phi) \right)^2
\hbox{ for } \phi \to \phi_\infty
\end{matrix} \right.
\hspace{10pt} ,
\end{equation}
which is in accordance with the general behavior given by
(\ref{Eq:Pot_limit_0}) and (\ref{Eq:V_limit_Lambda}).

\section{Comparison with cosmological observations}
\label{Sect_Observations}

%
\begin{figure}[ttt]
\begin{center}
\vspace*{-10pt}\begin{minipage}{8cm}
\includegraphics[width=8.0cm]{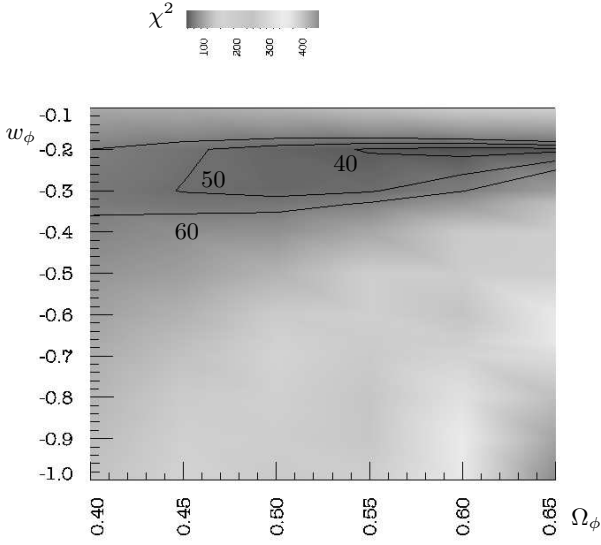}
\put(-20,62){$\Omega_\phi$}
\put(-234,208){$w_\phi$}
\put(-110,196){$40$}
\put(-160,190){$50$}
\put(-170,170){$60$}
\put(-180,252){$\chi^2$}
\end{minipage}
\vspace*{-70pt}
\end{center}
\caption{\label{Fig:Omega_tot_85_2d}
The $\chi^2$ values are shown in dependence on $w_\phi$ and $\Omega_\phi$
for $\Omega_{\hbox{\scriptsize tot}} = 0.85$
using RadPack for 41 data points.
The curves with $\chi^2=40$, 50 and 60 are indicated.
The minimum $\chi^2_{\hbox{\scriptsize min}} = 36$ occurs at
$\Omega_\phi=0.6$ and $w_\phi=-0.2$.
}
\end{figure}
\begin{figure}[ttt]
\begin{center}
\vspace*{-10pt}\begin{minipage}{8cm}
\includegraphics[width=8.0cm]{psplots/DX_2d_O_tot_90_bw.epsf}
\put(-20,62){$\Omega_\phi$}
\put(-234,208){$w_\phi$}
\put(-120,178){$40$}
\put(-158,165){$50$}
\put(-170,145){$60$}
\put(-180,252){$\chi^2$}
\end{minipage}
\vspace*{-70pt}
\end{center}
\caption{\label{Fig:Omega_tot_90_2d}
The $\chi^2$ values are shown in dependence on $w_\phi$ and $\Omega_\phi$
for $\Omega_{\hbox{\scriptsize tot}} = 0.9$
using RadPack for 41 data points.
The curves with $\chi^2=40$, 50 and 60 are indicated.
The minimum $\chi^2_{\hbox{\scriptsize min}} = 33$ occurs at
$\Omega_\phi=0.65$ and $w_\phi=-0.3$.
}
\end{figure}
\begin{figure}[ttt]
\begin{center}
\vspace*{-10pt}\begin{minipage}{8cm}
\includegraphics[width=8.0cm]{psplots/DX_2d_O_tot_95_bw.epsf}
\put(-20,62){$\Omega_\phi$}
\put(-234,208){$w_\phi$}
\put(-120,145){$40$}
\put(-160,142){$50$}
\put(-175,163){$60$}
\put(-180,252){$\chi^2$}
\end{minipage}
\vspace*{-70pt}
\end{center}
\caption{\label{Fig:Omega_tot_95_2d}
The $\chi^2$ values are shown in dependence on $w_\phi$ and $\Omega_\phi$
for $\Omega_{\hbox{\scriptsize tot}} = 0.95$
using RadPack for 41 data points.
The curves with $\chi^2=40$, 50 and 60 are indicated.
The minimum $\chi^2_{\hbox{\scriptsize min}} = 35$ occurs at
$\Omega_\phi=0.7$ and $w_\phi=-0.4$.
}
\end{figure}
\begin{figure}[ttt]
\begin{center}
\vspace*{-10pt}\begin{minipage}{8cm}
\includegraphics[width=8.0cm]{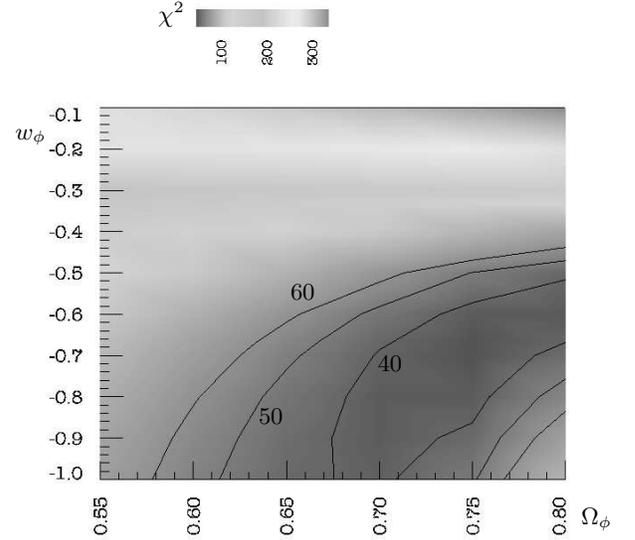}
\put(-20,62){$\Omega_\phi$}
\put(-234,208){$w_\phi$}
\put(-97,120){$40$}
\put(-142,100){$50$}
\put(-130,147){$60$}
\put(-180,252){$\chi^2$}
\end{minipage}
\vspace*{-70pt}
\end{center}
\caption{\label{Fig:Omega_tot_1_2d}
The $\chi^2$ values are shown in dependence on $w_\phi$ and $\Omega_\phi$
for $\Omega_{\hbox{\scriptsize tot}} = 1.0$
using RadPack for 41 data points.
The curves with $\chi^2=40$, 50 and 60 are indicated.
The minimum $\chi^2_{\hbox{\scriptsize min}} = 34$ occurs at
$\Omega_\phi=0.75$ and $w_\phi=-0.7$.
}
\end{figure}

After having discussed various aspects of quintessence models with a
constant equation of state $w_\phi$,
let us come to a detailed comparison with the CMB observations.
(A preliminary announcement of our results can be found in
\cite{Steiner_Aurich_2002}.)
We use in our comparison the following priors.
The Hubble constant $H_0 = h \times 100 \hbox{ km s}^{-1} \hbox{Mpc}^{-1}$
is set to $h=0.65$, and $\Omega_{\hbox{\scriptsize b}}=0.05$
(i.\,e.\ $\Omega_{\hbox{\scriptsize b}} h^2 = 0.021$)
is chosen in agreement with the current Big-Bang nucleosynthesis constraints.

The CMB anisotropy is computed according to
\cite{Ma_Bertschinger_1995,Hu_1998} using the conformal Newtonian gauge.
The relativistic components
are photons and three massless neutrino families
with standard thermal history.
For the photons, the polarization dependence on the Thomson cross section
is taken into account.
The recombination history of the universe is computed using RECFAST
\cite{Seager_Sasselov_Scott_1999}.
The non-relativistic components are baryonic and cold dark matter.
The initial conditions are given by an initial curvature perturbation
with no initial entropy perturbations in relativistic and
non-relativistic components.
Furthermore, we assume that there are no tensor mode contributions.
The initial curvature perturbation is assumed to be scale-invariant
which is ``naturally'' suggested by inflationary models.
The quintessence fluctuations are initially set to zero.
Other choices for the quintessence inhomogeneity would yield practically the
same results because these models are insensitive to the
initial conditions on the quintessence fluctuations
\cite{Dave_Caldwell_Steinhardt_2002}.

The CMB anisotropy of these quintessence models is compared with the
angular power spectrum $\delta T_l = \sqrt{l(l+1)C_l/2\pi}$
obtained by the experiments BOOMERanG \cite{Netterfield_et_al_2001},
MAXIMA-1 \cite{Lee_et_al_2001}, and DASI \cite{Halverson_et_al_2001}.
This corresponds to 41 data points.
The amplitude of the initial curvature perturbation is fitted
such that the value of $\chi^2$ is minimized with respect to 
these three experiments, where $\chi^2$ is computed using
RADPACK\footnote{See RADPACK homepage: \\
http://bubba.ucdavis.edu/$\sim$knox/radpack.html}.

\begin{figure}
\begin{center}
\begin{minipage}{8cm}
\includegraphics[width=8.0cm,height=5.0cm]{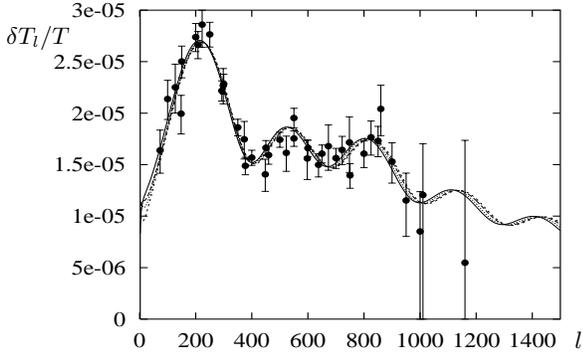}
\put(-10,5){$l$}
\put(-225,120){$\delta T_l/T$}
\end{minipage}
\vspace*{-15pt}
\end{center}
\caption{\label{Fig:aps_best_w_const}
The angular power spectrum $\delta T_l/T$ is presented for the
four best models for the cases
$\Omega_{\hbox{\scriptsize tot}} = 0.85$,
$\Omega_{\hbox{\scriptsize tot}} = 0.9$,
$\Omega_{\hbox{\scriptsize tot}} = 0.95$, and
$\Omega_{\hbox{\scriptsize tot}} = 1.0$.
These are determined by the minimum of $\chi^2$ shown
in figures \ref{Fig:Omega_tot_85_2d} to \ref{Fig:Omega_tot_1_2d}.
The data are taken from
\cite{Netterfield_et_al_2001,Lee_et_al_2001,Halverson_et_al_2001}.
}
\end{figure}

\begin{figure}[ttt]
\begin{center}
\vspace*{-20pt}\begin{minipage}{8cm}
\vspace*{-10pt}\includegraphics[width=9.0cm]{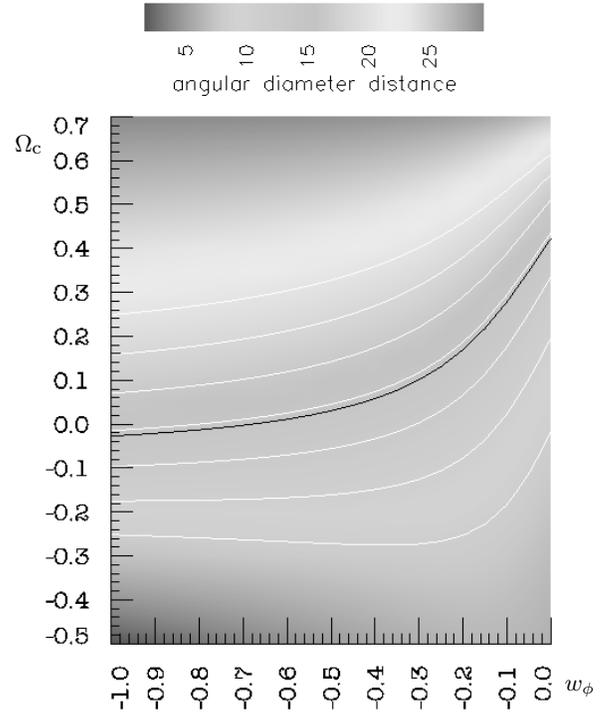}
\put(-40,48){$w_\phi$}
\put(-248,252){$\Omega_{\hbox{\scriptsize c}}$}
\end{minipage}
\vspace*{-45pt}
\end{center}
\caption{\label{Fig:angular_diameter_distance}
The angular-diameter distance $d_A$ is shown for models with
$\Omega_{\hbox{\scriptsize m}}=0.25$ in dependence on
$w_\phi$ and $\Omega_{\hbox{\scriptsize c}}$.
The curves show the contours of constant $d_A$.
Our models with the smallest $\chi^2$ value lie close to the black curve.
($z_{\hbox{\scriptsize sls}}=1100$ is assumed.)
}
\end{figure}

The figures \ref{Fig:Omega_tot_85_2d} to \ref{Fig:Omega_tot_1_2d} present
the values of $\chi^2$ for the four cases
$\Omega_{\hbox{\scriptsize tot}} = 0.85$,
$\Omega_{\hbox{\scriptsize tot}} = 0.9$,
$\Omega_{\hbox{\scriptsize tot}} = 0.95$, and
$\Omega_{\hbox{\scriptsize tot}} = 1.0$, respectively.
The values of $\chi^2$ are shown in dependence on $w_\phi$ and
$\Omega_\phi$.
Since $\Omega_{\hbox{\scriptsize tot}}$ is held fixed,
the matter density $\Omega_{\hbox{\scriptsize m}}$ is given by
$\Omega_{\hbox{\scriptsize m}} = \Omega_{\hbox{\scriptsize tot}} -
\Omega_\phi$ neglecting the small radiation contribution.
One observes that a decreasing $\Omega_{\hbox{\scriptsize tot}}$
demands an increasing $w_\phi$, i.\,e.\ a less negative value;
from $\Omega_{\hbox{\scriptsize tot}} = 0.85$ to
$\Omega_{\hbox{\scriptsize tot}} = 1.0$ the equation of state
shifts from $w_\phi=-0.2$ to $w_\phi=-0.7$.
Furthermore, all acceptable models require a dominant dark energy component
$\Omega_\phi\gtrsim 0.6$ and possess
$\Omega_{\hbox{\scriptsize m}} = 0.25$.
Since the minimum of $\chi^2$ is for all four cases of the
same order, $\chi^2_{\hbox{\scriptsize min}} = 33\dots 36$,
no definite value for the curvature is singled out.
Thus we do not find convincing hints pointing towards vanishing curvature,
i.\,e.\ a flat universe.
The four best models corresponding to the cases
$\Omega_{\hbox{\scriptsize tot}} = 0.85$,
$\Omega_{\hbox{\scriptsize tot}} = 0.9$,
$\Omega_{\hbox{\scriptsize tot}} = 0.95$, and
$\Omega_{\hbox{\scriptsize tot}} = 1.0$, respectively,
are nearly degenerated with respect to their angular power spectrum
$\delta T_l/T$ as can be seen in figure \ref{Fig:aps_best_w_const}.
With the current observational accuracy, one cannot discriminate between
these models.
A degeneracy between $\Omega_\phi$ and $w_\phi$ exists already in
the flat case, see e.\,g.\ 
\cite{Hu_Eisenstein_Tegmark_White_1999,Bean_Melchiorri_2002,%
Melchiorri_Mersini_Odman_Trodden_2002}.
If the assumption of flatness is dropped, a degeneracy
with respect to $\Omega_{\hbox{\scriptsize c}}$, $\Omega_\phi$ and
$w_\phi$ arises.
A special case, the degeneracy with respect to $\Omega_\Lambda$ and
the curvature $\Omega_{\hbox{\scriptsize c}}$ is discussed in
\cite{Efstathiou_Bond_1999}.
The full $(\Omega_{\hbox{\scriptsize c}}, \Omega_\phi, w_\phi)$-degeneracy 
can be inferred from \cite{Huterer_Turner_2001} in the
neighborhood of a special flat model,
where the shift $\Delta l_1$ of the first peak (see their Eq.\,(18))
from a flat $\Lambda$CDM model with $\Omega_{\hbox{\scriptsize m}} =0.3$
is analyzed.
This geometrical degeneracy arises through the angular-diameter distance
to the surface of last scattering having redshift
$z_{\hbox{\scriptsize sls}}$,
\begin{equation}
d_A \, = \, \label{Eq:angular_diameter_distance}
\frac{1}{H_0\sqrt{|\Omega_{\hbox{\scriptsize c}}|}
(z_{\hbox{\scriptsize sls}}+1)}
S_k\left( \sqrt{|\Omega_{\hbox{\scriptsize c}}|}
\int_{\frac 1{1+z_{\hbox{\tiny sls}}}}^1 \frac{dy}{\sqrt{g(y)}} \right)
\hspace{3pt} ,
\end{equation}
where $S_k(x)$ is $\sinh(x)$, $x$,  $\sin(x)$ for $k=-1,0,+1$, respectively,
if we include also the flat and the positive curvature case.
The models with the lowest values of $\chi^2$ possess nearly the same
angular-diameter distance $d_A$ to the surface of last scattering.
Furthermore, the models which for fixed $\Omega_{\hbox{\scriptsize c}}$ 
match the anisotropy, have all the same $\Omega_{\hbox{\scriptsize m}}=0.25$.
In figure \ref{Fig:angular_diameter_distance}
we show the angular-diameter distance $d_A$ to the
surface of last scattering for this value of $\Omega_{\hbox{\scriptsize m}}$.
Our models with the smallest $\chi^2$ lie close to the black marked contour
of constant $d_A$.
One infers from figure \ref{Fig:angular_diameter_distance}
that to the model with $\Omega_{\hbox{\scriptsize c}}=0$ and
$w_\phi=-0.7$ correspond models with the same $d_A$ having
$\Omega_{\hbox{\scriptsize c}}=0.05, 0.1, 0.15$ and
$w_\phi\simeq -0.43, -0.3, -0.22$, respectively.
Thus, this geometrical degeneracy allows also universes of
negative curvature.

One observes in figure \ref{Fig:angular_diameter_distance} that the
contours of constant $d_A$ have a larger slope for larger $w_\phi$,
i.\,e.\ for the models with more negative curvature.
This leads to more sharply confined regions of small values of $\chi^2$
in the parameter space with increasing $\Omega_{\hbox{\scriptsize c}}$,
i.\,e.\ decreasing $\Omega_{\hbox{\scriptsize tot}}$.
This trend is clearly visible in figures
\ref{Fig:Omega_tot_85_2d} to \ref{Fig:Omega_tot_1_2d}.

\begin{figure}
\begin{center}
\begin{minipage}{8cm}
\includegraphics[width=8.0cm,height=5.0cm]{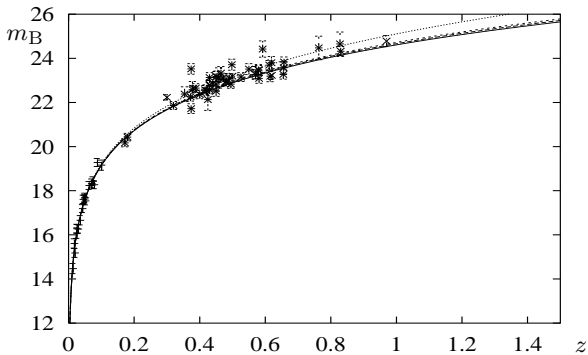}
\put(-10,5){$z$}
\put(-225,123){$m_{\hbox{\scriptsize B}}$}
\end{minipage}
\vspace*{-15pt}
\end{center}
\caption{\label{Fig:SN_best_w_const}
The magnitude-redshift relation $m_{\hbox{\scriptsize B}}(z)$
is shown for the four best models possessing
$\Omega_{\hbox{\scriptsize tot}} = 0.85$,
$\Omega_{\hbox{\scriptsize tot}} = 0.9$,
$\Omega_{\hbox{\scriptsize tot}} = 0.95$, and
$\Omega_{\hbox{\scriptsize tot}} = 1.0$
in comparison with the supernovae Ia data
\cite{Hamuy_et_al_1996,Riess_et_al_1998,Perlmutter_et_al_1999}.
}
\end{figure}

In figure \ref{Fig:SN_best_w_const} the magnitude-redshift relation
$m_{\hbox{\scriptsize B}}(z)$ is shown for the four best models.
One observes a very similar behavior in the three cases with negative
curvature consistent with the supernovae Ia observations
\cite{Hamuy_et_al_1996,Riess_et_al_1998,Perlmutter_et_al_1999}.
The upper dotted curve belonging to the flat case gives a slightly
better fit than the curves belonging to the cases with negative curvature
but the latter are nevertheless consistent with these observations.
At this point, it is important to be aware of the ongoing discussions
whether the extinction in the host galaxies is indeed negligible
for high $z$ supernovae \cite{Sullivan_et_al_2002}
or whether this indicates an inconsistent treatment of host galaxy extinction
when the extinction is only taken into account for low $z$ supernovae
\cite{Rowan-Robinson_2002,Farrah_Meikle_Clements_Rowan-Robinson_Mattila_2002}.
Furthermore, the hints for an accelerated expansion are also weakened
if  supernovae not observed before maximum light are excluded
from the analysis \cite{Rowan-Robinson_2002}.
Thus, it is probably save not to reject universes with negative curvature
which need only a $0.1-0.2$ mag shift in order to be in perfect agreement
with the observations.
At this point it should be noted, as e.\,g.\ emphasized recently in
\cite{Ellis_Stoeger_McEwan_Dunsby_2002}
that while inflation is taken to predict that the universe is very close
to flat, it does {\it not} imply that the spatial sections are
{\it exactly} flat.

\begin{figure}[ttt]
\begin{center}
\begin{minipage}{8cm}
\includegraphics[width=5.5cm,angle=270]{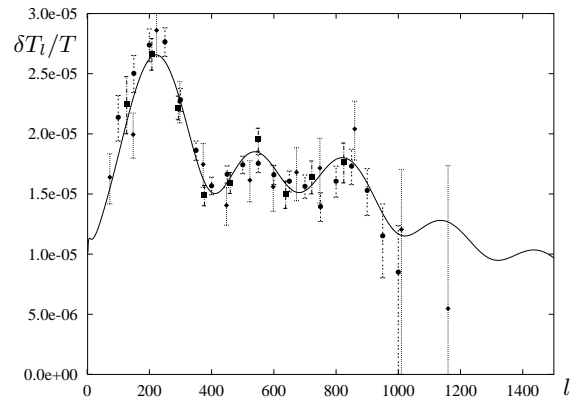}
\put(-7,-152){$l$}
\put(-215,-22){$\delta T_l/T$}
\end{minipage}
\vspace*{-20pt}
\end{center}
\caption{\label{Fig:APS}
The angular power spectrum $\delta T_l/T$ for the quintessence potential
(\ref{Eq:V_13}) for $w_\phi=-\frac 13$
in comparison with the experiments
\cite{Netterfield_et_al_2001,Lee_et_al_2001,Halverson_et_al_2001}
as discussed in the text.
}
\end{figure}

\begin{figure}
\begin{center}
\begin{minipage}{8cm}
\includegraphics[width=5.5cm,angle=270]{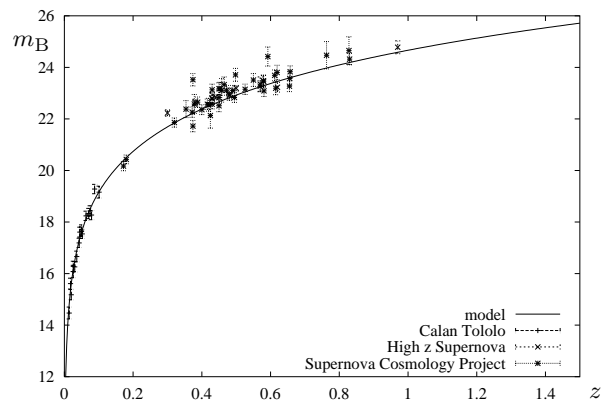}
\put(-7,-152){$z$}
\put(-225,-19){$m_{\hbox{\scriptsize B}}$}
\end{minipage}
\vspace*{-20pt}
\end{center}
\caption{\label{Fig:SN}
The $m_{\hbox{\scriptsize B}}(z)$ relation in comparison with the
supernovae Ia data
\cite{Hamuy_et_al_1996,Riess_et_al_1998,Perlmutter_et_al_1999}
for the quintessence potential (\ref{Eq:V_13}) for $w_\phi=-\frac 13$.
}
\end{figure}

Let us now turn to the special case $w_\phi = -\frac 13$,
i.\,e.\ to the potential (\ref{Eq:V_13})
derived in Sect.\,\ref{Sect_Analytic_Expressions_13}.
Here we choose the following values for the cosmological parameters:
$\Omega_{\hbox{\scriptsize cdm}} = 0.25$,
i.\,e.\ $\Omega_{\hbox{\scriptsize m}} = 0.30$;
$\Omega_\phi=0.60$, i.\,e.\ $\Omega_{\hbox{\scriptsize tot}} = 0.90$
or $\Omega_{\hbox{\scriptsize c}} = 0.10$,
which gives
$\Omega_{\hbox{\scriptsize tot}}(\infty) = \frac 67 \simeq 0.86$.
We then obtain $\eta_{\hbox{\scriptsize eq}} = 0.0087$;
$z_{\hbox{\scriptsize rec}} = 1089$,
i.\,e.\ $\eta_{\hbox{\scriptsize rec}} = 0.0198$;
$z_{\hbox{\scriptsize m}} = \sqrt{(\Omega_\phi+\Omega_{\hbox{\scriptsize c}})
/\Omega_{\hbox{\scriptsize r}}} - 1 \simeq 83$,
i.\,e.\ $\eta_{\hbox{\scriptsize m}} = 0.106$;
$z_{\hbox{\scriptsize m}\phi}=1$,
i.\,e.\ $\eta_{\hbox{\scriptsize m}\phi} = 0.687$;
$\eta_0 = 0.8938$ corresponding to an age of the universe of
$t_0 = 12.16 \hbox{ Gyr}$.
In Fig.\,\ref{Fig:Omega_eta} we show the density parameters 
$\Omega_{\hbox{\scriptsize x}}(\eta)$ as a function of conformal time $\eta$.
Fig.\,\ref{Fig:APS} shows the prediction of the model for the
angular power spectrum $\delta T_l = \sqrt{l(l+1)C_l/2\pi}$ of the CMB
anisotropy in comparison with the BOOMERanG \cite{Netterfield_et_al_2001},
MAXIMA-1 \cite{Lee_et_al_2001},
and DASI \cite{Halverson_et_al_2001} experiments.
One observes that the model describes the data very well
($\chi^2=35$ for $N_{\hbox{\scriptsize data}} =41$).
The first three acoustic peaks occur at $l_1=222$, $l_2=538$
and $l_3=820$, in excellent agreement with the observations.
In Fig.\,\ref{Fig:SN} we show the magnitude-redshift relation
$m_{\hbox{\scriptsize B}}(z)$ in comparison with the supernovae Ia data
\cite{Hamuy_et_al_1996,Riess_et_al_1998,Perlmutter_et_al_1999}.
Again we observe good agreement with the data.


To summarize, we conclude that the quintessence model with
$w_\phi=-\frac 13$ and $\Omega_{\hbox{\scriptsize tot}}=0.90$
is in excellent agreement with present observations.
We have thus demonstrated that it is too early to claim
that present data have already established that the universe is flat.
It remains to be seen whether future observations give additional support
to the idea that our universe is close to flat, but not exactly flat,
and that its spatial geometry is hyperbolic.

\bibliographystyle{apsrev}
\bibliography{../bib_astro} 

\end{document}